\documentclass{aa}
\usepackage{txfonts}
\usepackage{graphicx}
\begin{document}
   \title{New signposts of massive star formation in the S235A-B region}

   \author{Marcello Felli\inst{1},
 Fabrizio Massi\inst{1}, Massimo Robberto\inst{2}, 
and  Riccardo Cesaroni\inst{1}
          }

   \offprints{F. Massi, fmassi@arcetri.astro.it}

   \institute{INAF-Osservatorio Astrofisico di Arcetri,
              Largo E. Fermi 5, I-50125 Firenze, Italy 
\and
Space Telescope Science Institute, 3700 San Martin Drive, Baltimore, MD 21218, USA
}

   \date{Received; accepted }

\abstract{
We report on new aspects of the star-forming region S235AB revealed
through high-resolution observations at radio and
mid-infrared wavelengths. 
Using the Very Large Array,
we carried out sensitive observations of S235AB
in the cm continuum (6, 3.6, 1.3, and 0.7) and in
the 22 GHz water maser line. These were complemented with Spitzer
Space Telescope Infrared Array Camera 
archive data to clarify the correspondence between radio and IR sources.
We made also use of newly presented data from the Medicina water
maser patrol, started in 1987, to study the variability of the water masers
found in the region.
S235A is
a classical HII region whose structure is now well resolved. To the south,
no radio continuum emission is detected either from the compact molecular core 
or from the jet-like structure observed at 3.3 mm, suggesting emission from dust
in both cases. We find two new compact radio continuum  sources
(VLA-1 and VLA-2) and three separate maser spots.
VLA-1 coincides with one of the maser spots and with a previously
identified IR source (M1). VLA-2 lies towards S235B and represents
the first radio detection from this peculiar nebula 
that may represent an ionized wind from a more evolved star.
The two other maser spots coincide
with an elongated structure previously observed within the molecular
core in the C$^{34}$S line.
This structure is perpendicular to a bipolar molecular outflow
observed in HCO$^{+}$(1-0) and may trace the associated equatorial disk.
The Spitzer images reveal
a red
object towards the molecular core. This is the most viable candidate for the
embedded source originating the outflow and maser phenomenology.
The picture emerging from these and previous data
shows the extreme complexity of a small ($\leq 0.5$~pc)
star-forming region where widely different stages of stellar evolution
are present. 

}

   \authorrunning{M. Felli et al.}
   \titlerunning{The S235A-B star-forming region}

\maketitle
%

\section{Introduction}

   \begin{figure*}
   \centering
   \includegraphics[angle=-90,width=16cm]{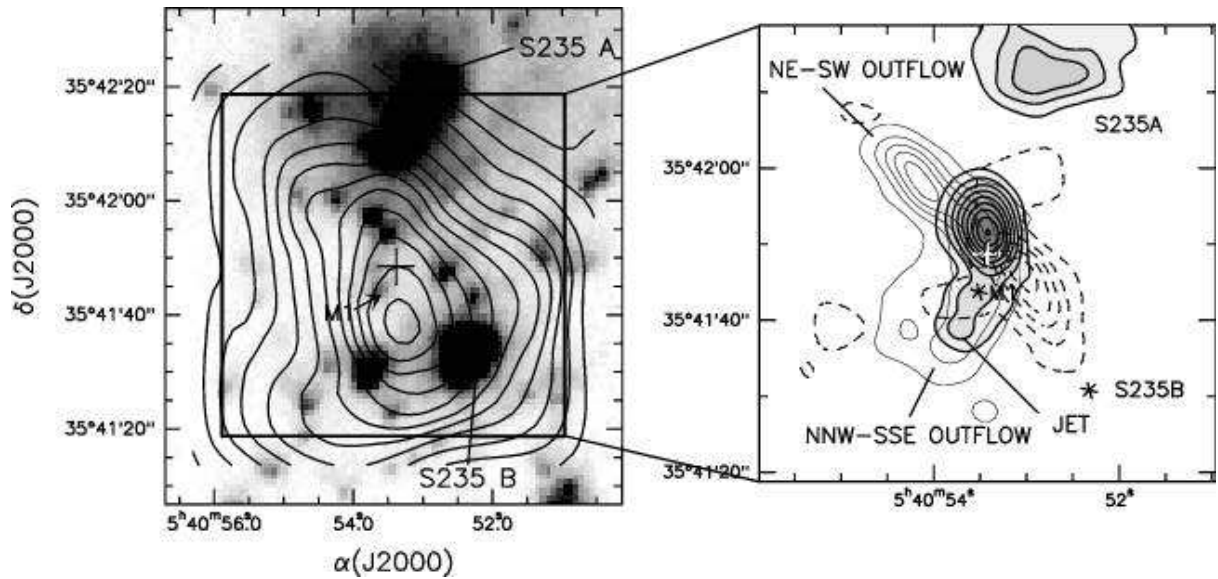}
    \caption{Overview of the S235A-B star forming complex
%
    before the observations presented in this paper.
    Left: overlay of the single-dish HCN map (contours) from Cesaroni et
    al. (1999) with the $K$-band image of the S235A-B region
    from Felli et al. (1997). Right: overlay of the 
    interferometric maps of the 3.3 mm continuum emission (thick contours and 
    grey scale) with the HCO$^{+}$(1--0)
    outflows (thin solid line: $V < -20.9$ km s$^{-1}$,
    dashed line: $V> -15.5$ km s$^{-1}$) from Felli et al.\ (2004). 
    In both, a cross marks the location of the H$_{2}$O maser at -61.2 km 
    s$^{-1}$ from Tofani et al. (1995). The infrared
    source with the largest near-IR excess detected in previous works, M1, 
    is also indicated.
%
    The offsets between M1, the water maser, the HCN peak, and the mm 
    continuum peak are all larger than the position uncertainties.
         \label{Fig0}}
   \end{figure*}
%
This paper continues the study of the star-forming complex 
S235A-B (see Felli et al.\ 2004 and references therein), focussing in 
particular  on a  
deeply embedded Young Stellar Object (YSO) found 
between S235A and S235B, close to a water maser.
The presence of the YSO is implied from the typical signposts of early stellar
evolution, including two molecular outflows, a hot molecular core,  a 
sub-millimeter
peak, and a water maser. The YSO represents the youngest object in the  
star-forming complex. 

The site  morphology
%
derived from all previous observations
is summarised in  Fig.~\ref{Fig0}.
S235A is a small optical nebulosity coinciding with a compact, but 
well-resolved, HII region. It appears as a less-evolved region, with respect 
to the more extended and diffuse HII region S235 
located further north, and probably is unrelated to the S235A-B complex.
S235A lies at the northern edge of a molecular clump 
%
that
represents
the brightest peak of a more extended molecular cloud (Evans \& Blair 1981;
Nakano \& Yoshida 1986; Cesaroni et al.\ 1999). 
   
   Lying $\sim40\arcsec$ south of S235A (0.35 pc at the assumed distance of
1.8 kpc; Nakano \& Yoshida 1986), at the SW edge of the molecular
core, S235B is a smaller diffuse nebulosity detected both in the optical and 
near-IR, exhibiting a  near-IR excess and
intense emission in optical and IR hydrogen lines
(H$\alpha$, Br$\gamma$), but without a radio continuum counterpart, 
making it 
%
a rather
peculiar object. 
   In H$\alpha$, it consists of an unresolved peak superimposed
on a circular nebula that is $\sim10\arcsec$ in diameter (Krassner et al.\ 1982;
Alvarez et al.\ 2004). S235B appears to be a young star with an expanding 
ionized envelope surrounded by a diffuse nebulosity (Felli
et al.\ 1997).

A large-scale molecular outflow was first found in $^{12}$CO(1--0) by  
Nakano \& Yoshida (1986) with a resolution of $\sim14\arcsec$; it was 
centred at S235B and aligned in a NE-SW direction, about 35$\arcsec$ 
(0.3 pc) in 
length. 
Felli et al.\ (1997)
confirmed the blue lobe of the outflow in $^{13}$CO(2--1) with a
resolution of  $\sim11\arcsec$, but failed to detect the red lobe.
   
Near-IR images revealed a highly obscured stellar cluster
between S235A and S235B, i.e. centred on the water maser,
with several sources with IR excess 
(Felli et al.\ 1997), in particular source M1, which exhibits the largest  near-IR 
excess. This source has been suggested to be the 
candidate YSO supplying energy to the $-60$ km s$^{-1}$ water maser,
but with great uncertainty since it lies more than 5$\arcsec$ to the south 
(see also Fig.~\ref{Fig0}).

The above picture was clarified 
by  Felli et al.\ (2004), who presented  high-resolution (between 
2$\arcsec$ and 4$\arcsec$) mm line (HCO$^{+}$, C$^{34}$S, H$_{2}$CS, 
SO$_{2}$, and CH$_{3}$CN) and continuum observations, together with far-IR
observations. 

   A compact molecular core (hereafter, the mm core) was found 
both in the mm continuum (hot dust emission, $T_{\rm dust} = 
T_{\rm CH_{3}CN}\sim30$ K) and in the 
molecular lines, peaking close to the water maser position 
and well-separated from S235A and S235B. 
Two molecular outflows were found in HCO$^{+}$(1--0)
centred on the mm core. One of them (hereafter, the NE-SW outflow) is 
aligned  along the same NE-SW direction of the large-scale outflow detected by
Nakano \& Yoshida (1986). It spans $\sim0.4$ pc,
and has an estimated mass of 9 M$_{\sun}$ and a mechanical luminosity 
$>19$ L$_{\sun}$.
The other (hereafter, the NNW-SSE outflow) is more compact and aligned in a
NNW-SSE direction. It  spans $\sim0.3$ pc and has
an estimated mass of 4 M$_{\sun}$ and a mechanical luminosity of
$\sim0.3$ L$_{\sun}$. 

Felli et al.\ (2004) derived an upper limit of
$\sim10^{3}$ L$_{\sun}$ for the bolometric luminosity of the mm core and
suggested the presence of an embedded intermediate-mass YSO 
driving the NE-SW outflow and supplying the energy for the 
$-60$ km s$^{-1}$ water maser. 
Studying the high velocity red and blue emission of C$^{34}$S(5--4)
towards the mm core, they also found a compact structure with a 
velocity gradient perpendicular to
the NE-SW outflow that might represent the signature of a circumstellar disk.

An elongated structure (called a ``jet'') protruding from the mm core and  
coinciding with the blue lobe of the NNW-SSE outflow 
was also detected in the continuum at 3.3 mm (see Fig.~\ref{Fig0}, right) 
and, with a lower signal-to-noise ratio, also  at 1.2 mm. 
The spectral index of the jet, $\alpha \sim 0.6$ (defined as 
S$_{\nu}\propto \nu^{\alpha}$) 
is rather uncertain, but different from that of the mm core, 
$\alpha \sim 2.5$, suggesting that the emission 
might arise in an ionized wind rather than being due to dust. 
Radio jets 
are often observed at the base of molecular outflows (Rodr\'{\i}guez
1997; Anglada et al.\ 1998; Beltr\'{a}n et al.\ 2001) in both low-mass 
and high-mass
star-forming regions (e.\ g., Rodr\'{\i}guez 1996). They are characterized 
by spectral indices in the range $-0.1$ to $\sim1$ and are elongated in 
the outflow direction.
Expanding ionized envelopes also have a spectral index 
$\alpha$ = 0.6 (Panagia \& Felli 1975). 

Water masers are one of the most reliable signposts of early phases 
in star formation 
(see e.g. Tofani et al.\ 1995) since they provide the best indication
of the position of the required powering source, i.e. the YSO. 
They occur both in low-mass 
(see, e.g., Furuya et al.\ 2001, 2003) and in high-mass (see Churchwell
2002, and references therein) star-forming regions and are often found 
to be associated with outflowing matter. 
Sometimes they are also found in close association with radio jets 
(e.g.\ G\'{o}mez et al.\ 1995). 

%
The presence of a water maser in this region had been known since the
observations of  Henkel et al.\ (1986) and  Comoretto et al.\ (1990),
but with insufficient spatial resolution to properly locate it in the
region. Only with Very Large Array (VLA) cm line 
observations (Tofani et al.\ 1995)
and interferometric mm observations of the continuum sources
(Felli et al.\ 2004), was the location of the water maser
in-between S235A and S235B, almost coincident with the mm core, 
firmly established.  This proved that the water maser  is  not 
associated with  either of the two nebulosities and
that  a local early type star, presumably the YSO within the mm core,
is needed for its excitation.
In the VLA observations, maser emission was only searched for in a
limited velocity range, around $-60$ km s$^{-1}$, since at the time 
this  was the only component detected by single-dish observations.

The water maser in S235A-B has been monitored with the Medicina 
radio telescope since 1987,
with coverage $\sim4$ times per year since 1993. 
These observations
%
revealed
at least three separate velocity components: the one already known 
at $\sim -60$ km s$^{-1}$, one between $-20$ to
$-30$ km s$^{-1}$, and one between $-10$ to 10 km s$^{-1}$. The last two
are always very weak (max 10-20 Jy) and  exhibit 
strong variations. 
Whether they are all related to the same YSO or to 
%
separate
ones could not be established  from single-dish observations because of 
%
low
spatial resolution.  This drove us to carry out new VLA line observations 
at 22 GHz covering the whole maser velocity 
interval. All three velocity components found in the Medicina 
%
data
were active at the time of this VLA observation.

   The S235A-B region contains other masers, namely methanol
(CH$_{3}$OH)   and SiO (Nakano \& Yoshida 1986; Haschick et al.\ 1990; 
Harju et al.\ 1998).  Kurtz et al.\
(2004) included S235A-B in their recent VLA survey of the CH$_{3}$OH maser
line at 44 GHz.  This is a class I methanol 
%
source,
and it is believed to trace outflow activity. These authors found
a cluster of 6 CH$_{3}$OH masers spread over an area of 
%
a
few square arcsec around the water maser spot at 
   $\sim -60$ km s$^{-1}$. 

%
To
clarify the nature of the embedded YSO and its relation with 
%
the outflow found
in the S235A-B region, we have performed an
extensive observational program using the VLA and the Medicina radio telescopes,
complemented by archival Spitzer data.

The two primary goals  of the new VLA continuum observations
were: 1) to clarify the nature of the ``jet'' and to derive its spectrum over
a larger frequency interval and 2) to search further for cm emission from
ionized hydrogen in the mm core.  VLA observations in the water maser line 
together with the Medicina patrol can give indications on the
location and activity of the maser in the star-forming region.
Finally, for a better understanding of the precise correspondence 
between 
%
IR and radio sources,
in particular the precise role of M1 and the
possibility of detecting IR emission from the mm core,
archive Spitzer-IRAC observations of the S235A-B region in the four
wavelengths (3.6, 4.5, 5.8, and 8 $\mu$m) were retrieved and
analyzed.

In Sect.~\ref{obs} we describe the observations, and in Sect.~\ref{res}
we present the results, while in Sect.~\ref{discuss} we discuss our findings
and how they enrich our current understanding of the S235A-B region.
%
In Sect.~\ref{conclu}, the main results are summarised. The reader
can refer to Fig.~\ref{schema} for a comprehensive sketch of the
S235A-B star forming-region, including the latest data.

\section{Observations and data reduction}
\label{obs}

%
\begin{table*}
\begin{minipage}{\columnwidth}
\caption{Summary of VLA observations.
\label{obs:tab}}     
\centering          
\renewcommand{\footnoterule}{}
\begin{tabular}{c c c c c c c c}     
\hline\hline
Date & Frequency & \multicolumn{2}{c}{Synthesized beam\footnote{natural 
weighting}} & 
%
Largest
& Int.\ time & rms & Notes \\ 
 &  & Size & PA & 
%
Angular Scale
 & & & \\
     & (GHz) & (arcsec $\times$ arcsec) & (degree) & 
%
(arcsec)
& (sec) & (mJy/beam) & \\
\hline
07/03/2004, 26/02/2004 & 4.75 & $4.8 \times 4.3$ & $-8.2$ & 300 & 1200 & 0.09 & \\ 
07/03/2004 & 8.45 & $2.8 \times 2.6$ & $-20.8$ & 180 & 1800 & 0.08 & \\ 
26/02/2004 & 23 & $1 \times 1$ &  $-40$ & 60 & 7224 & 0.03 & cont. \\
07/03/2004 & 23 & $1 \times 0.9$ &  $-83.7$ & 60 & 434 & 0.05 & line \\
07/03/2004 & 45 & $0.6 \times 0.6$ & $88.6$ & 43 & 17590\footnote{fast switching} & 0.08 & \\
\hline
\end{tabular}
\end{minipage}
\end{table*}
%

\subsection{VLA observations}

The observations carried out with the VLA of the
National Radio Astronomy Observatory 
(NRAO)\footnote{The National Radio Astronomy 
Observatory is a facility of the National Science Foundation operated 
under cooperative agreement by Associated Universities, Inc.} were made with the VLA in the C configuration 
on February 26 and  on March 7, 2004.
The location of the water maser at $-60$ km s$^{-1}$ was used as phase centre; 
its coordinates are:\\
$\alpha(2000)=05^{h}40^{m}53.42^{s}$, 
$\delta(2000)=35\degr41\arcmin48\farcs 8$.\\
The observations 
%
consisted of
calibrator-target-calibrator scans; in the
Q-band (0.7 cm) the fast-switching mode was used. 
The phase calibrator was 0555+398, with 
boot-strapped fluxes of 5.3 Jy (6 cm), 4.55 Jy (3.6 cm), 2.71 Jy (1.3 cm),
and 1.85 Jy (0.7 cm); the absolute amplitude calibrator was 3C147 
(0542+498),
with assumed fluxes of 7.9 Jy (6 cm), 4.8 Jy (3.6 cm), 1.8 Jy (1.3 cm),
and 0.9 Jy (0.7 cm). The pointing accuracy was checked on 0555+398 (at 3.6 cm)
every hour. Line observations at 1.3 cm were carried out with 
the correlator
set to 64 channels and a bandwidth of $6.25$ MHz. This resulted in a velocity 
resolution of 1.3 km
s$^{-1}$ over a range of 84 km s$^{-1}$, sufficient to cover the whole 
velocity range
of the water masers. The receivers were tuned so as to have the band centred
at $-30$ km s$^{-1}$ with respect to the Local Standard of Rest.
A summary of the observations is contained in Table~\ref{obs:tab}.

Since the  forthcoming analysis is based on a comparison of the position of the 
radio sources with those
of  near- and mid-IR sources, we checked 
whether the radio
coordinate system is consistent with the near-IR coordinate system. 
To this end, we searched the 2MASS point source catalogue
for a near-IR counterpart of our phase calibrator, finding an object at
the same position as the calibrator within 0$\farcs$1. 
This  gives us confidence that the two coordinate systems are consistent 
with each other within this limit.

\subsection{Medicina observations}
The single-dish Medicina 
radio telescope\footnote{The Medicina VLBI radio telescope is operated
by the Radioastronomy Institute of INAF, Italy} (HPBW 1\farcm 9) 
observations are part 
of a monitoring project of
a large sample of water masers. An autocorrelator with 1024 channels
and a 10 MHz bandwidth is usually employed. The typical
sensitivity for a 5-minute integration is of the order of 1 Jy, and the
average calibration error is about 20\%.
For a more detailed description of the radio telescope and the 
relevant parameters of the 
water masers patrolling,  we refer to Valdettaro et al.\ (2002) and 
Brand et al.\ (2003).

\subsection{Spitzer-IRAC observations}

Spitzer-IRAC observations of a large area around the S235A-B region in the four
wavelengths (3.6, 4.5, 5.8, and 8 $\mu$m) were extracted from the
Spitzer public archive.
The observations are part of the GTO program 201 ``The Role of
Photodissociation Regions in High Mass Star
Formation'' (Principal Investigator G. Fazio). Integration time is 12s at
all filters.

The positional accuracy is better than $1\arcsec$.
Point-source FWHM resolutions range from $\sim1\farcs 6$ at 3.6 $\mu$m
to $\sim1\farcs 9$ at 8.0 $\mu$m.
%
The IRAC bands are large and may contain various features,
depending on the environment being observed. Among these, the
most important for our case are polycyclic aromatic
hydrocarbon (PAH) features (3.6, 5.8, and 8.0 bands) and the
Br$\alpha$ line (4.5 band).
An overview of IRAC is given by Fazio et al.\ (2004a).

\section{Results}
\label{res}

\subsection{Spitzer-IRAC observations}

\begin{figure*}
   \centering
   \includegraphics[angle=90,width=16cm]{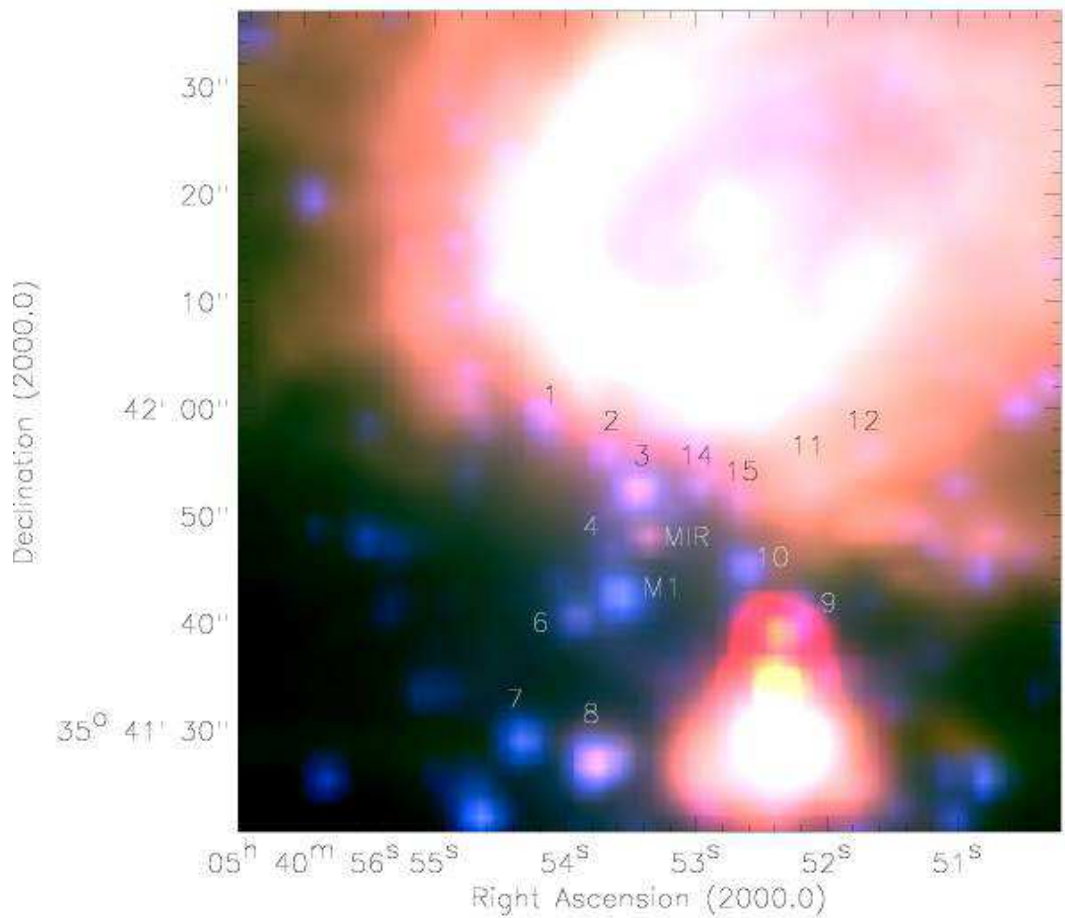}
    \caption{ Three colour (4.5 blue, 5.8 green, and 8.0 $\mu$m red) image of the
S235A-B region. The labels indicate the position of
M1, S235AB-MIR, and the other stars of the cluster for which we 
performed photometry.         
\label{spitzercolor}}
   \end{figure*}

A colour coded image of the region covering S235A-B, obtained by combining
4.5, 5.8 and 8.0 $\mu$m observations,
is shown in Fig.~\ref{spitzercolor}.

The two diffuse nebulosities S235A and S235B clearly dominate the 
extended emission.

\begin{figure}
   \centering
   \includegraphics[angle=-90,width=8cm]{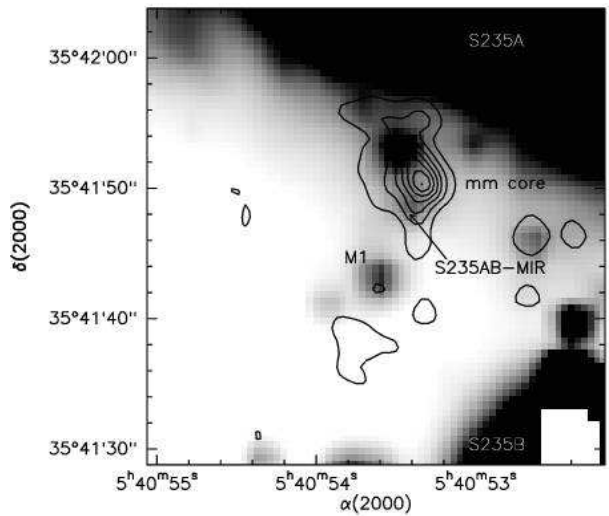}
      \caption{The 1.2  mm core (contours) 
from Felli et al.\ (2004) overlaid on the 
       5.8  $\mu$m Spitzer-IRAC image (grey scale, S235A and S235B have been  
saturated to show the weak emission from S235AB-MIR). 
The positions of the mm core, 
S235AB-MIR, and  M1 are indicated. 
         \label{redsource}}
   \end{figure}

Most of the point sources  detected in the $K$-band and shown in 
Fig.~\ref{Fig0} are also present here.
Two 
%
sources deserve more attention
in view of the present work:
1) M1, 
%
which
is detected at all bands, and 2) 
%
a source
(hereafter, S235AB-MIR) 
%
detected only at 4.5, 5.8, and 8.0 $\mu$m, which
is present in the area of the mm core. Both 
sources are marked in Figs.~\ref{spitzercolor} and 
~\ref{redsource}  where
we show an overlay of the mm core (at 1.2 mm) with
the 8.0 $\mu$m IRAC image. S235AB-MIR lies $1\farcs 5$ to the 
south of the peak of the  mm core.

   \begin{figure}
   \centering
   \includegraphics[angle=-90,width=8cm]{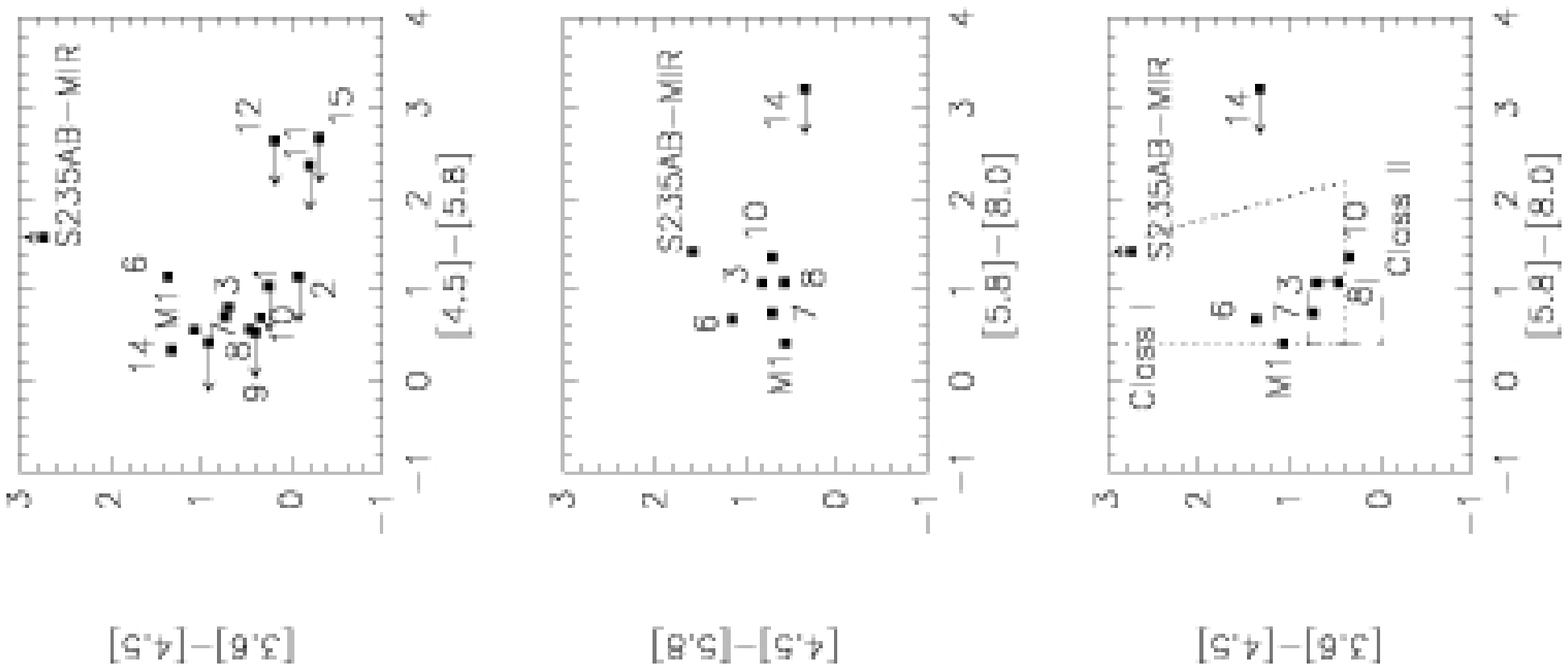}
      \caption{Colour-colour plots of the most relevant sources in the
area of the S235A-B cluster. The identifying numbers and symbols
are the same as those used in Table~\ref{spitz:tab} and Fig.~\ref{spitzercolor}.
In the bottom box, the regions occupied by Class I
and Class II sources are enclosed with dotted and dashed lines, respectively.
%
When only an upper limit to the flux density could be estimated in one 
of the bands, the corresponding point in the plot is marked by an arrow. 
         \label{colcol}}
   \end{figure}

We performed aperture photometry on the IRAC sources
found near to the mm-core by using DAOPHOT in IRAF.
For all four bands, we selected
a radius of 2 pix ($\sim1$ FWHM; 1 pix $\sim1\farcs 2$) 
and a 2-pix wide annulus with an
inner radius of 4 pix, to account for the highly
variable background. We applied aperture corrections as estimated
from the IRAC PSFs retrieved from the Spitzer Web Page
(http://ssc.spitzer.caltech.edu/obs/). To derive the
photometric zero points,
we used the zero-magnitude fluxes given by Fazio et al.\ (2004b).
                                                                                
Photometry was done only on 
the most relevant sources present in the area of the S235A-B cluster,
which are  indicated in Fig.~\ref{spitzercolor}.
Positions and flux densities in the four IRAC bands of the sources
labelled in Fig.~\ref{spitzercolor} are given in 
Table~\ref{spitz:tab}

%
\begin{table*}
\caption{MIR fluxes for the Spitzer-IRAC sources towards the mm core.
\label{spitz:tab}}
\renewcommand{\footnoterule}{}
\begin{tabular}{l c c c c c c}
\hline\hline
  ID  & \multicolumn{2}{c}{Position} &   $F_{\nu}(3.6)$    &
      $F_{\nu}(4.5)$   &   $F_{\nu}(5.8)$     &    $F_{\nu}(8.0)$   \\
      & $\alpha(2000)$ & $\delta(2000)$ & (mJy)   &  (mJy)  &  (mJy)
      &   (mJy)  \\
\hline
   1  & $05^{h}40^{m}54.2^{s}$ & $35\degr41\arcmin59\arcsec$ &
         $8.6 \pm 0.8$  &  $7.0 \pm 0.8$    & $<12$ & $<39$ \\
   2  & $05^{h}40^{m}53.7^{s}$ & $35\degr41\arcmin56\arcsec$ &
         $7.8 \pm 0.9$  &  $4.7 \pm 1.2$ & $<11$ & $<19$ \\
   3  & $05^{h}40^{m}53.5^{s}$ & $35\degr41\arcmin53\arcsec$ &
        $17.2 \pm 0.3$  & $21.2 \pm 0.6$ &  $29 \pm 1$ &  $43 \pm 4$   \\
   4  & $05^{h}40^{m}53.7^{s}$ & $35\degr41\arcmin47\arcsec$ &
        $1.9 \pm 0.2$  &  $2.9 \pm 0.4$  & $<3$ & $<5$   \\
   5 (M1) & $05^{h}40^{m}53.6^{s}$ & $35\degr41\arcmin43\arcsec$ &
        $8.6 \pm 0.1$  & $15.0 \pm 0.1$ &  $16 \pm 1$ &  $13 \pm 1$  \\
   6  & $05^{h}40^{m}54.0^{s}$ & $35\degr41\arcmin41\arcsec$ &
        $2.1 \pm 0.1$  &  $4.8 \pm 0.1$ &   $9 \pm 1$ &   $9 \pm 1$    \\
   7  & $05^{h}40^{m}54.4^{s}$ & $35\degr41\arcmin29\arcsec$ &
        $5.9 \pm 0.1$  &  $7.5 \pm 0.1$ &   $9 \pm 1$  &  $10 \pm 1$   \\
   8  & $05^{h}40^{m}53.8^{s}$ & $35\degr41\arcmin27\arcsec$ &
       $33.6 \pm 0.1$  & $33.1 \pm 0.1$ &  $36 \pm 1$ &  $54 \pm 1$    \\
  9$^a$  & $05^{h}40^{m}52.2^{s}$ & $35\degr41\arcmin41\arcsec$ &
        $5.0 \pm 0.3$    &  $4.7 \pm 0.2$ & $<5$ & $<24$  \\
  10  & $05^{h}40^{m}52.7^{s}$ & $35\degr41\arcmin45\arcsec$ &
        $8.5 \pm 0.2$  &  $7.6 \pm 0.2$ &   $9 \pm 1$ &  $18 \pm 7$     \\
  11  & $05^{h}40^{m}52.2^{s}$ & $35\degr41\arcmin55\arcsec$ &
        $5.4 \pm 0.8$  &  $2.9 \pm 0.7$ & $<17$ & $<95$  \\
  12  & $05^{h}40^{m}51.7^{s}$ & $35\degr41\arcmin57\arcsec$ &
        $3.5 \pm 0.8$  &  $2.7 \pm 0.6$ & $<20$ & $<59$   \\
  13 (S235AB-MIR) & $05^{h}40^{m}53.4^{s}$ & $35\degr41\arcmin49\arcsec$ &
       $<0.6$ &   $5.0 \pm 0.3$ &  $14 \pm 1$ &  $28 \pm 2$ \\
  14  & $05^{h}40^{m}53.0^{s}$ & $35\degr41\arcmin53\arcsec$ &
        $1.7 \pm 0.9$  &  $3.7 \pm 1.0$  &  $3 \pm 4$ & $<35$      \\
  15  & $05^{h}40^{m}52.7^{s}$ & $35\degr41\arcmin52\arcsec$ &
        $2.4 \pm 1.1$  &  $1.2 \pm 0.9$ & $<9$ & $<22$   \\
\hline
\end{tabular}

\vspace*{1mm}
$^a$ N-E of a close by source only visible in the $5.8$ $\mu$m band.
\end{table*}
%

Colour-colour plots are shown in Fig.~\ref{colcol}. In the 
[5.8]--[8.0]/[3.6]--[4.5] plot, the region occupied by Class I and Class II 
objects (Allen et al.\ 2004) is indicated. 
M1 has colours typical of a Class I object, while S235AB-MIR 
is located in 
%
a part of the plot redward of that
occupied by YSOs of mass $\sim6$ M$_{\sun}$, 
according to Whitney et al.\ (2004), i.e. the region corresponding to
[5.8]--[8.0] $\ge 1$ and [3.6]--[4.5] $\ge 1$.
 
\subsection{Radio continuum}
\subsubsection{S235A}
   \begin{figure}
   \centering
   \includegraphics[angle=0,width=8cm]{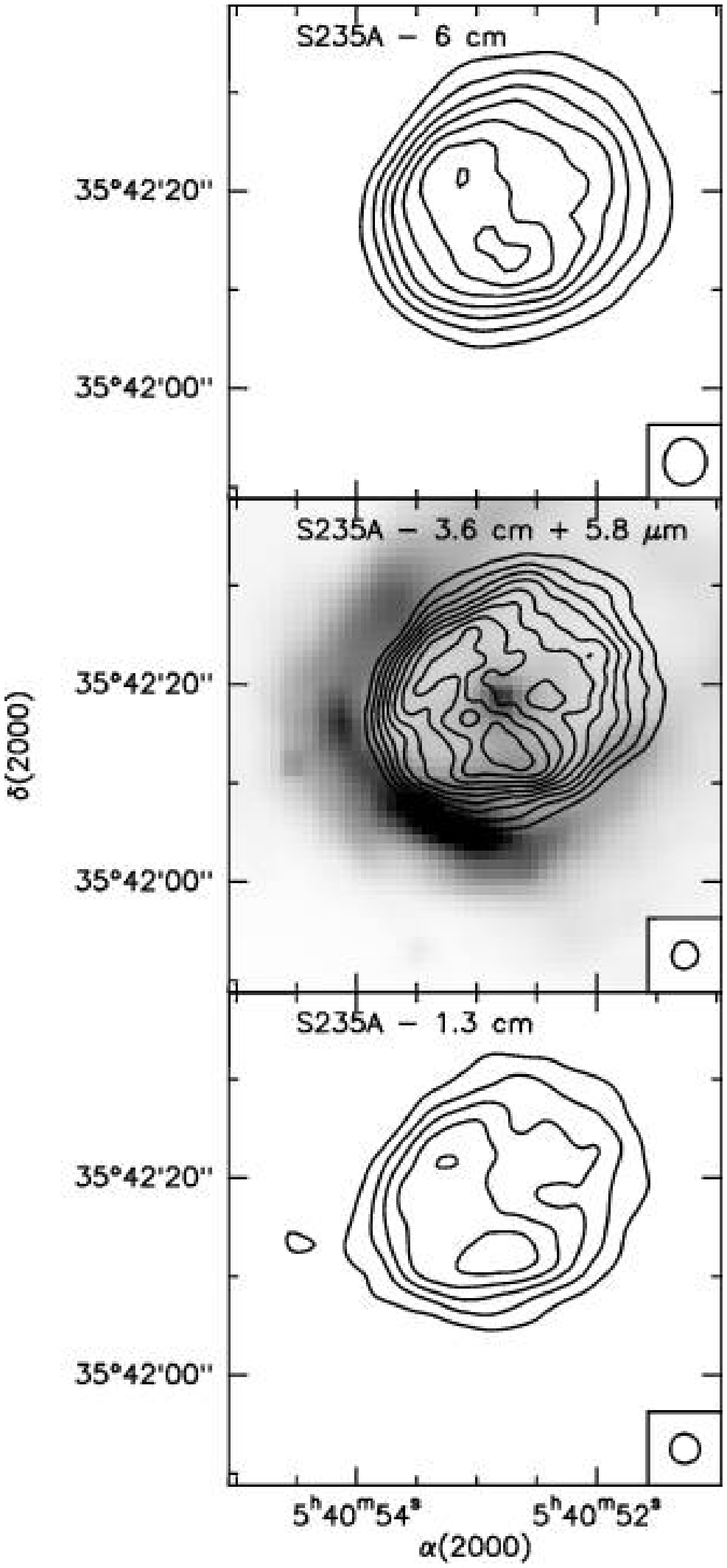}
      \caption{Maps of the continuum radio emission from S235A at 6, 3.6, and
1.3 cm (contours).  Levels are from 2 to 14 mJy/beam by 
        2 mJy/beam for 6 cm, from 1 to
        5 mJy/beam by 0.5 mJy/beam for 3.6 cm, and
	0.1 to 0.5 mJy/beam by 0.1 mJy/beam for 1.3 cm. 
The 1.3 cm map has been smoothed to a lower resolution to increase
the S/N ratio.  The  3.6 cm  map is
       overlaid with the 5.8 $\mu$m Spitzer-IRAC image (grey scale).
         \label{s235a:radio}}
   \end{figure}
%

The VLA radio continuum maps 
are dominated by the emission from the 
compact HII region S235A.
We calculated the integrated flux at all frequencies 
over the area within the 3$\sigma$
contour of the emission at 6 cm (i.e. at the lowest resolution); 
the obtained values are
listed in Table~\ref{cont:tab}. The integrated flux at 6 cm is in very 
good agreement
with previous measurements, yielding $270 \pm 40$ mJy. Israel \& Felli (1978)
found $220 \pm 30$ mJy at 21 cm and suggested a partially thick emission 
at 6 cm, but the
ratio of fluxes at 6 and 3.6 cm that we derive from our data is in agreement 
with  an optically thin emission. The presence of an ionizing star of
spectral type B0.5 derived from the radio fluxes in previous
works (e.g. Felli et al.\ 1997) is therefore  confirmed.
In Fig.~\ref{s235a:radio}, we show the  maps of S235A at 6, 3.6, and 1.3 cm. 
The 3.6 cm map is overlaid on 
the 5.8 $\mu$m Spitzer-IRAC image and
shows a well-resolved spherical shape. The Spitzer-IRAC
image reveals the ionizing star S235A* at the centre of the nebula, previously 
detected in the  K band (Felli et al.\ 1997).
The main feature of the radio maps  is the asymmetry of the isophotal contours in the 
SE-NW direction. In the IRAC images, the morphology clearly indicates 
the presence of a brighter ridge SE of S235A*.

The contours at 1.3 cm outline the brightest parts of the radio ridge. 
In the original maps, the emission is more fragmented because of the 
high resolution and low surface brightness. 
In Fig.~\ref{s235a:radio}, the map has been smoothed to a 
%
resolution of $3 \arcsec$.
At 1.3 and 0.7 cm, the radio fluxes are lower than expected 
from an optically thin emission, as was also found by Felli et al.\ (2004) at 3.3 mm. 
We attribute this to an instrumental effect caused by the  filtering
of extended structures in the interferometric observations
%
(see Table~\ref{obs:tab}).

\subsubsection{The mm core, the jet, and the radio compact sources VLA-1 and VLA-2}

At none of the 4 VLA wavelengths 
%
were we
able to 
detect emission from the mm core where the presence of an intermediate-mass YSO 
is suggested by the mm observations. 
This excludes any thermal emission from a UCHII region
associated with the YSO with a flux density above the noise level
given in Table~\ref{cont:tab}  and implies
%
that emission from the core is dominated by dust.
At the same time, our non-detection is not in contradiction 
with the existence of 
%
a dust core.
In fact, 
extrapolating at 0.7 cm, the flux density measured at 3.3 mm (20 mJy)
with a spectral index of 2.5
(Felli et al.\ 2004), we obtain  3 mJy for the flux expected from the core.
Felli et al.\ (2004) estimate that the core diameter in the continuum
at the highest resolution (1.2 mm) is $\sim3\arcsec$; hence, assuming that
all the emission is uniformly distributed in a circle of $1\farcs 5$ in radius
and using the synthesized beam size at 0.7 cm given in Table~\ref{obs:tab},
we obtain  an expected  flux of $\sim0.12$ mJy/beam, i.\ e.\ $< 2\sigma$
(see Table~\ref{obs:tab}). Hence, dust emission at 0.7 cm 
could be present  below the sensitivity limit of our observations.

Similarly, no extended emission  
from  the jet was observed at any of our four frequencies. 
While this could be an effect of over-resolution
and low surface brightness of the jet  at the shortest wavelengths, 
it definitely rules out the hypothesis
of an ionized jet at 6 and 3.6 cm. In fact, at 6 cm, the flux density 
per beam area
extrapolated  from the 3.3 mm flux in the hypothesis of an ionized 
jet (i.e. using a spectral index $\alpha$ = 0.6) would be a factor of 3 higher 
than the upper limit quoted in Table~\ref{cont:tab}.

\begin{figure}
   \centering
   \includegraphics[angle=-90,width=8cm]{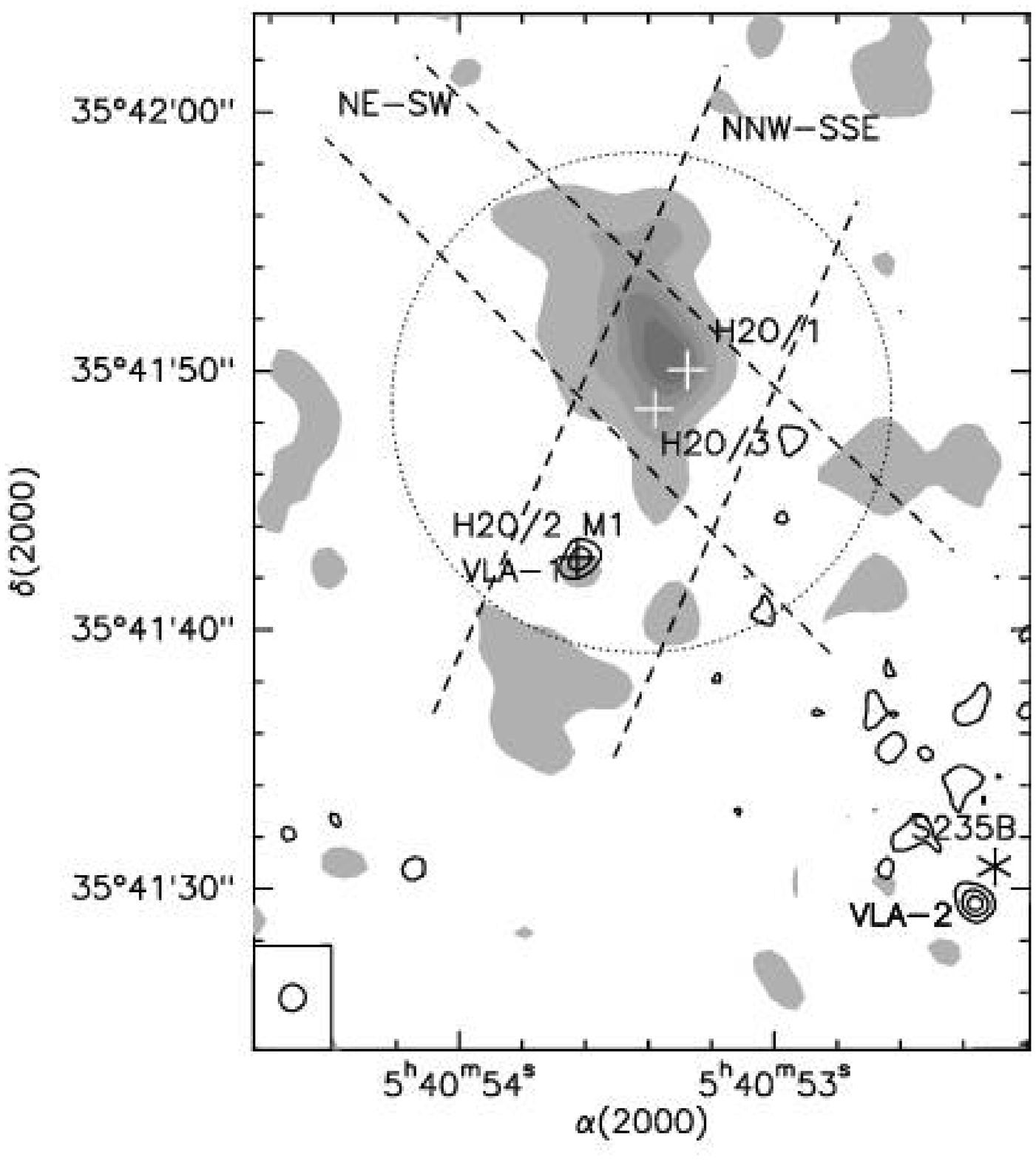}
      \caption{Map of the continuum radio emission at 1.3 cm  of the area
around the water masers (contours)
       overlaid with the map of the continuum 1.2 mm emission 
       (grey scale) from Felli et al.\ (2004). 
Levels are from 0.12 ($\sim4\sigma$) to
        0.48 mJy/beam in steps of 0.12 mJy/beam. The locations of the
	three H$_{2}$O maser spots detected in this work are indicated by ``+''.
	M1 and the centre of S235B (asterisk) are also indicated. 
The dashed lines define 
	the directions of the two outflows.
	The dotted circle is the primary beam at 1.2 mm.
	The synthesized beam at 1.3 cm is drawn in the bottom left-hand corner.
         \label{Fig1}}
   \end{figure}

At 1.3 and 0.7 cm, where the resolution is higher, 
two nearly unresolved sources are present.
They have been named VLA-1 and VLA-2 and are indicated in Fig.~\ref{Fig1},
where we show the VLA
map at 1.3 cm (contours) overlaid on the Plateau de Bure map at 1.2 mm 
(grey scale) from Felli et al.\ (2004). 
We derived the integrated fluxes of the two  sources within the $3\sigma$ level
on the maps obtained with natural weighting.
At 3.6 and 6 cm, the two sources fall within the sidelobes of 
S235A. This results in a noise higher than that predicted from the 
total integration time (see Table~\ref{obs:tab}). 
Using different weightings 
to partially filter out the extended emission does not yield any significant 
improvement in the
measurements. The flux densities are listed in Table~\ref{cont:tab}; the 
upper limits at 6 cm (for both) and at 3.6 cm (for VLA-2)  refer
to a point source.

%

VLA-1 lies $\sim10\arcsec$
south of the mm core and is  located within the jet, 
almost along its axis (see Figs.~\ref{Fig0} and 
\ref{Fig1}). 
It coincides with a small component of the fragmented  jet observed at 1.2 mm
and, most noticeably, with the K-band source M1.

VLA-2 lies  within the boundary of the 
S235B nebulosity. VLA-2  represents 
the first radio detection of this peculiar region. 
Our flux densities of $\sim0.5$ mJy (see Table~\ref{obs:tab})
are not in conflict with the previously derived  upper limits of 5 mJy at 6 cm 
($10\arcsec$ beam; Israel \& Felli 1978)
and 0.3 mJy at 3.6 cm ($0\farcs 1$ beam; Tofani et al.\ 1995). 

We have checked  the probability that VLA-1 and VLA-2 are 
background  sources. The expected number of extra\-galactic sources
at 1.3 cm in the field of view of Fig.~\ref{Fig1} ($\sim1$ square arcmin) 
based on Eq.~(A11) of Anglada et al.\ (1998) is $N\sim 0.05$, making this
possibility very unlikely.

%
\begin{table*}
\begin{minipage}{\columnwidth}
\caption{Radio continuum fluxes. Upper limits are $3 \sigma$ noise levels.
\label{cont:tab}}     
\centering          
\renewcommand{\footnoterule}{}
\begin{tabular}{l c c c c c c}     
\hline\hline
Source & \multicolumn{2}{c}{Position} & \multicolumn{4}{c}{$S_{\nu}$(mJy)} \\ 
     & $\alpha(2000)$ & $\delta(2000)$ & 0.7 cm & 1.3 cm & 3.6 cm & 6 cm \\
\hline
S235A & $05^{h}40^{m}52.70^{s}$ & $35\degr42\arcmin21\arcsec$ & $114 \pm 3$\footnote{Corrected
	for primary beam attenuation.} &
	$172 \pm 1$ & $248 \pm 2$ & $257 \pm 7$ \\
VLA-1 & $05^{h}40^{m}53.60^{s}$ & $35\degr41\arcmin43\arcsec$ & $0.59 \pm 0.08$ & $0.44 \pm 0.03$ & $0.39 \pm 0.08$ & $< 0.27$ \\
VLA-2 & $05^{h}40^{m}52.40^{s}$ & $35\degr41\arcmin30\arcsec$ & $0.48 \pm 0.08$ & $0.47 \pm 0.03$ & $<0.24$ & $<0.27$ \\ 
\hline
\end{tabular}
\end{minipage}
\end{table*}
%

\subsection{H$_{2}$O masers}

\subsubsection{VLA observations}

In the selected velocity range (roughly from $-70$ to 10 km s$^{-1}$), 
three maser spots were detected 
above the  $5\sigma$ noise.
We have determined the flux densities and positions of the maser spots
by 2-dimensional Gaussian fits in each channel. All velocity components
in the  same spot are  spatially unresolved.
Coordinates  of the three maser spots and
flux density for each velocity peak are listed in Table~\ref{h2o:tab}
their relations to the other features present in the area are shown
in Figs.~\ref{Fig1} and \ref{Figmaser}.

\begin{figure}
   \centering
   \includegraphics[angle=-90,width=8cm]{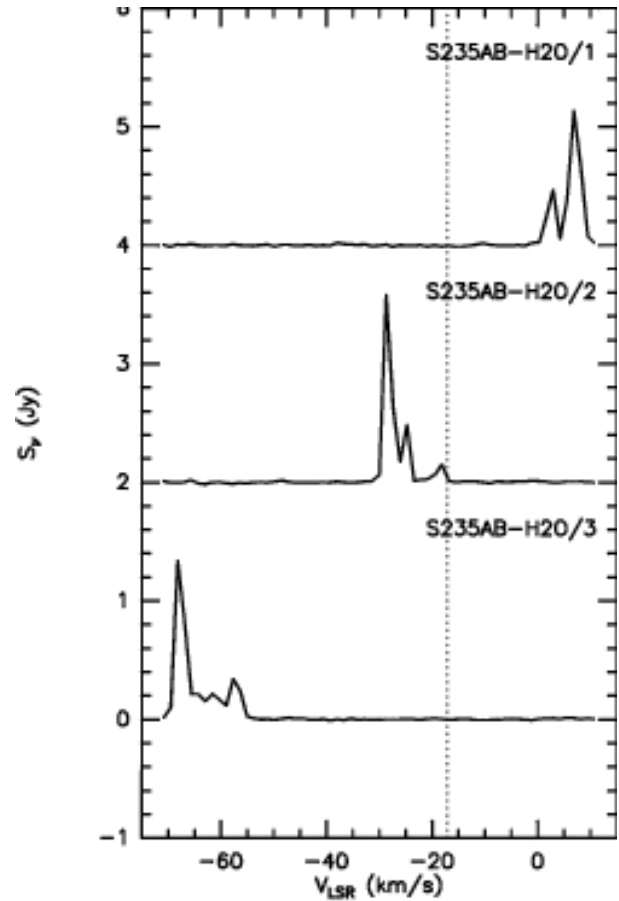}
      \caption{H$_{2}$O VLA spectra averaged on a circle 1$\arcsec$ in radius
      and  centred on the locations  of each of the 3 maser spots. The name 
      of the spot is indicated above the corresponding
      spectrum. The intensity scale is offset by 2 and 4 Jy for clearness
of the display. The vertical line defines the velocity of the molecular
cloud.
         \label{Fig2}}
   \end{figure}

It is important to note that the three maser spots 
cover different velocity ranges, as shown in Fig.~\ref{Fig2}, so that
there is no velocity overlap among the three spatial components.

One of them (S235A-B-H2O/3)
coincides, within $\sim0\farcs 5$, with that found with the VLA at 
$\sim -60$ km s$^{-1}$ by Tofani et al.\ (1995) and emits in the 
same velocity range.
The other two (S235A-B-H2O/1 and S235A-B-H2O/2) occur at radial 
velocities that had not been searched for in the previous
VLA observation because at that time they were not detectable in single-dish
observations, but they have since been revealed in the Medicina patrol. 

The maser luminosity for each
spot was obtained  by integrating the line emission 
within the respective  velocity ranges.
The results are listed in Table~\ref{lumi:tab}, 
along with the corresponding velocity range.
The H$_{2}$O luminosity is typical of masers associated with 
far-Infrared (FIR) sources
of $10^{3} - 10^{4}$ L$_{\sun}$ (see Palagi et al.\ 1993). This is
consistent with the 
upper limit of the bolometric luminosity inferred for the mm core 
by Felli et al.\ (2004).

%
\begin{table}
\begin{minipage}{\columnwidth}
\caption{Water masers: fluxes and positions.
\label{h2o:tab}}     
\centering          
\renewcommand{\footnoterule}{}
\begin{tabular}{c c c c c}     
\hline\hline
Name & \multicolumn{2}{c}{Position}  & & \\ 
S235A-B & $\alpha(2000)$ & $\delta(2000)$ & $V_{\rm LSR}$ & $S_{\nu}$ \\ 
&                &                &  (km s$^{-1}$) & (Jy) \\             
\hline
H2O/1 & $05^{h}40^{m}53.27^{s}$ & $35\degr41\arcmin50\farcs 0$ & 7 & 1.20 \\ 
            &            &                             & 3 & 0.50 \\
H2O/2 & $05^{h}40^{m}53.63^{s}$ & $35\degr41\arcmin43\farcs 0$ & $-18$ & 0.16 \\
            &			&                             & $-25$ & 0.52 \\
            &			&                             & $-29$ & 1.66 \\
H2O/3 & $05^{h}40^{m}53.38^{s}$ & $35\degr41\arcmin48\farcs 6$ & $-58$ & 0.37 \\
            &                   &                             & $-62$ & 0.23 \\
	    &		        &                             & $-64$ & 0.22 \\
            &			&                             & $-68$ & 1.44 \\
\hline
\end{tabular}
\end{minipage}
\end{table}

From Figs.~\ref{Fig1}  and \ref{Figmaser}, 
S235A-B-H2O/2 clearly coincides with VLA-1 and M1
and does not seem to be 
directly related to the mm core.
The other two spots (S235A-B-H2O/1 and S235A-B-H2O/3)
are very  close to the mm core and are perpendicularly aligned
to the NE-SW outflow, $\sim2\arcsec$ apart from each other. 
%
The possibility that they might be tracing a disk or torus
perpendicular to the NE-SW outflow will be examined in Sect.~\ref{discuss}.

%
\begin{table}
\begin{minipage}{\columnwidth}
\caption{Water masers: line luminosity.
\label{lumi:tab}}     
\centering          
\renewcommand{\footnoterule}{}
\begin{tabular}{c c c}     
\hline\hline
Name & $L_{\rm H_{2}O}$\footnote{
%
All masers are assumed to be at the distance of the S235A-B complex.
} 
     & $\Delta V_{\rm LSR}$ \\ 
     & ($10^{-7}$ L$_{\sun}$) & (km s$^{-1}$) \\             
\hline
S235A-B-H2O/1 & $3.1$ & 9 to $-1$ \\
S235A-B-H2O/2 & $3.3$ & -17 to -30 \\
S235A-B-H2O/3 & $4.1$ & -55 to -71 \\
\hline
\end{tabular}
\end{minipage}
\end{table}
%

%

   \begin{figure}
   \centering
   \includegraphics[angle=-90,width=8.5cm]{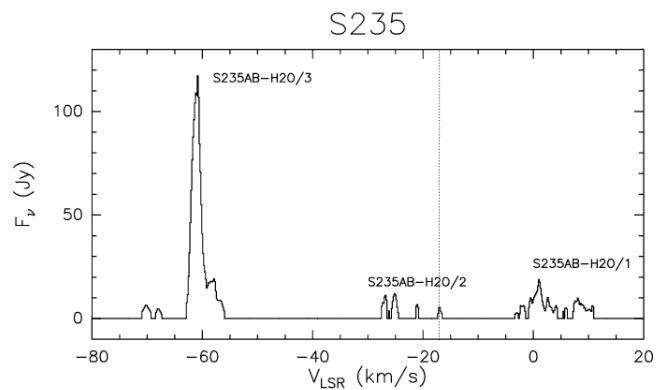}
      \caption{Upper envelope of all water maser spectra
      observed with the Medicina radio telescope towards S235A-B,
      until July 2005. The three corresponding  
      maser spots detected with the VLA
      observations are indicated. The vertical line defines the velocity of 
      the molecular cloud.
         \label{medi:maser}}
   \end{figure}
%

   \begin{figure*}
   \centering
   \includegraphics[angle=-90,width=16cm]{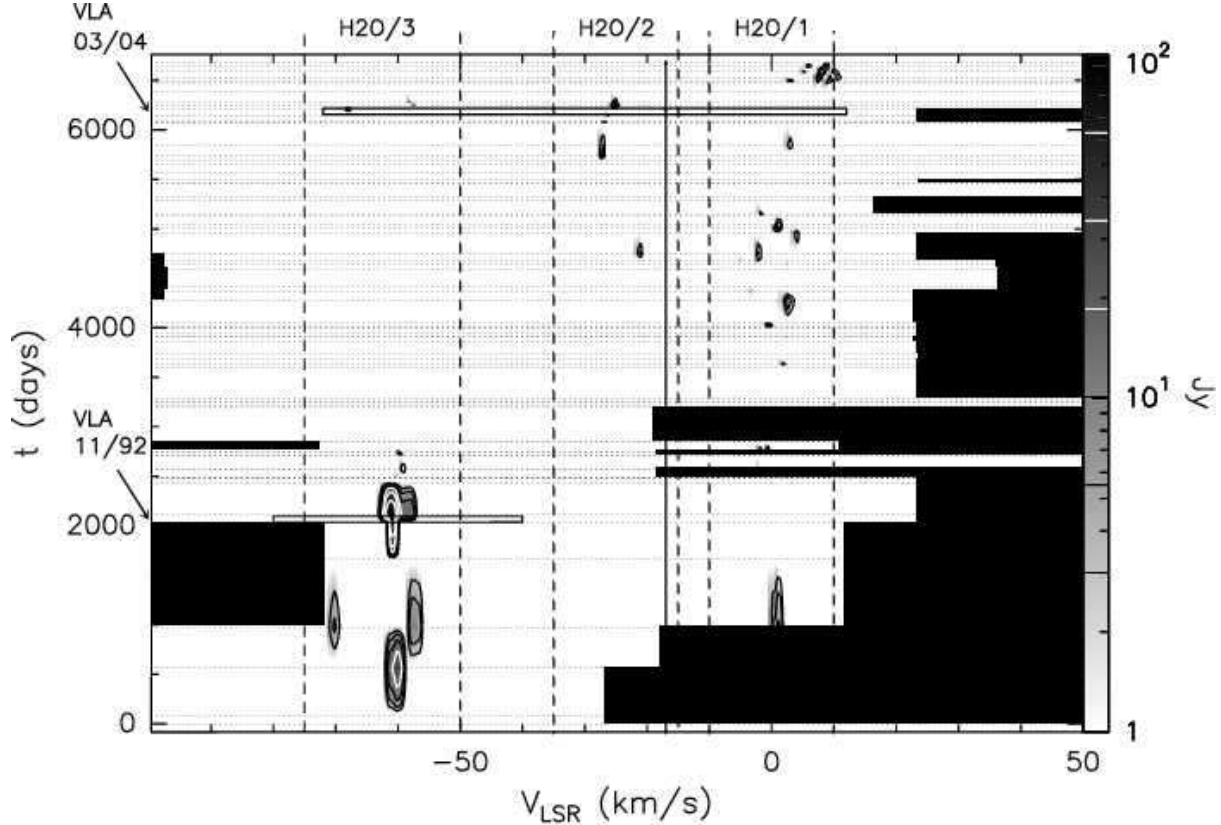}
    \caption{Time-velocity-intensity plot of the water maser emission from
S235A-B observed with the Medicina radio telescope. 
The starting date is  March 31, 1987.
The dates of the first VLA observation by Tofani et al.\ (1995) and of 
the present observations are indicated (month/year) with an arrow and 
%
their velocity ranges are
enclosed within a long rectangle. The three maser
spots are indicated by bracketing the velocity ranges with vertical dashed lines. 
The vertical solid line defines the velocity of the molecular
cloud. The black areas are time-velocity regions with no observations.
         \label{Fig5}}
   \end{figure*}
%

   \begin{figure*}
   \centering
   \includegraphics[angle=-90,width=16cm]{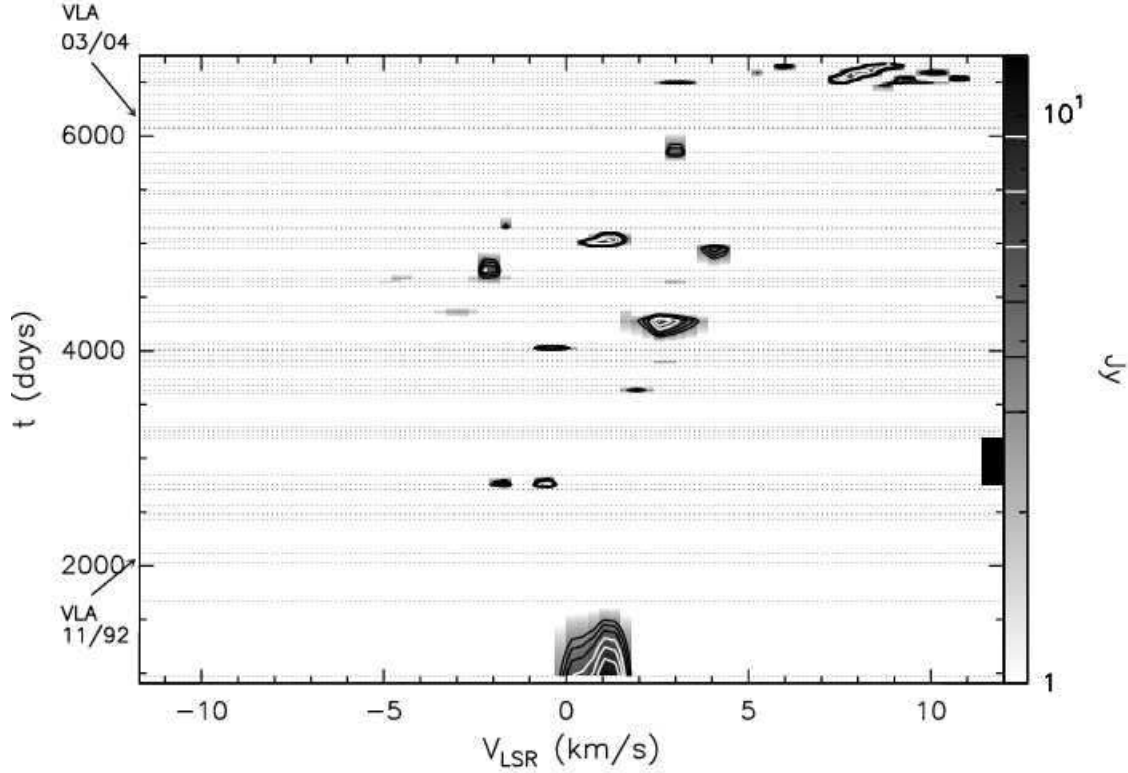}
    \caption{Time-velocity-intensity plot of the water maser emission from
S235A-B observed with the Medicina radio telescope over the 
velocity range from $-10$ to $+10$ km s$^{-1}$, which corresponds to 
the maser spot S235A-B-H2O/1. 
         \label{Fig6}}
   \end{figure*}
%

\subsubsection{Medicina observations}

The different velocity ranges of the maser spots detected
with the VLA 
make the low-resolution Medicina observations a useful tool to follow
the evolution of each spot, with no confusion arising from the fact that
they are all within the Medicina beam. The upper envelope of the water
maser emission using all the available single-dish observations is shown in 
Fig.~\ref{medi:maser}. When compared with Fig.~\ref{Fig2}, it shows
that the emissions from the three spatially separated VLA maser spots 
occur at very different velocity ranges, being thus separated (although spatially 
unresolved) in the Medicina observations, as well.

In Fig.~\ref{Fig5}, we show the time-velocity-intensity plot from the Medicina
patrol. An indicative value of the noise level in these observations throughout
the entire period is of the order of 1-2 Jy.
The starting date corresponds to March 31, 1987. The patrol is 
sparser at the beginning. After 1992, there are about 4 observations 
every year. Following the separation in velocity of the three maser spots,
each velocity component (at $\sim-60$, $\sim-25$, and $\sim0$ km s$^{-1}$)
is labelled in Fig.~\ref{Fig5} with the corresponding VLA name. 
The dates of the two VLA observations are indicated. 
The systemic velocity of the
molecular cloud  ($-17$ km s$^{-1}$) is also indicated.
Only S235A-B-H2O/2 occurs at a  
velocity close to that of the thermal molecular lines; the other two components
(S235A-B-H2O/1 and S235A-B-H2O/3) emit well outside the width of the 
thermal molecular lines.

The biggest flare occurred  from the component
S235A-B-H2O/3 at the time of the first VLA observation. The source
then disappeared below the noise and came up again just shortly before 
the second VLA observation. 

Component S235A-B-H2O/2 was undetectable  
for most of the time and appeared above the noise only after 2000.

Component S235A-B-H2O/1, although always rather weak, is present 
most of the time. Its most noteworthy aspect is that the velocity 
changes up to $\pm$ 5 km s$^{-1}$ around  a mean value of $\sim0$ km s$^{-1}$.
To better illustrate this effect, the time-velocity-intensity plot for the
velocity range from $-10$ to 10 
km s$^{-1}$ is shown in Fig.~\ref{Fig6}. 

To make sure that the change in velocity is a real effect and not an 
instrumental one and to provide an independent estimate of the accuracy
on the velocity, we have examined another water maser (G32.74--0.08), that is
also included in the Medicina patrol, and that was chosen because it is characterized 
by a single, narrow, and intense velocity component.
For this maser, the velocity of the peak displays a maximum deviation 
from the mean value over
the entire period of $< 0.1$ km s$^{-1}$, a factor of 50 smaller than the
velocity spread observed in S235A-B-H2O/1.

\section{Discussion}

\label{discuss}

\subsection{S235A}

In Fig.~\ref{s235a:radio}, the overlay of the 3.6 cm map
with the 5.8 $\mu$m Spitzer image shows that the peak of
the IR emission occurs in a shell outside the boundary of the radio
emission. This proves that PAH and  thermal dust   emissions are mostly
 located beyond the ionization front in the Photodissociation Region.
                                                                                
   \begin{figure}
   \centering
   \includegraphics[angle=-90,width=8cm]{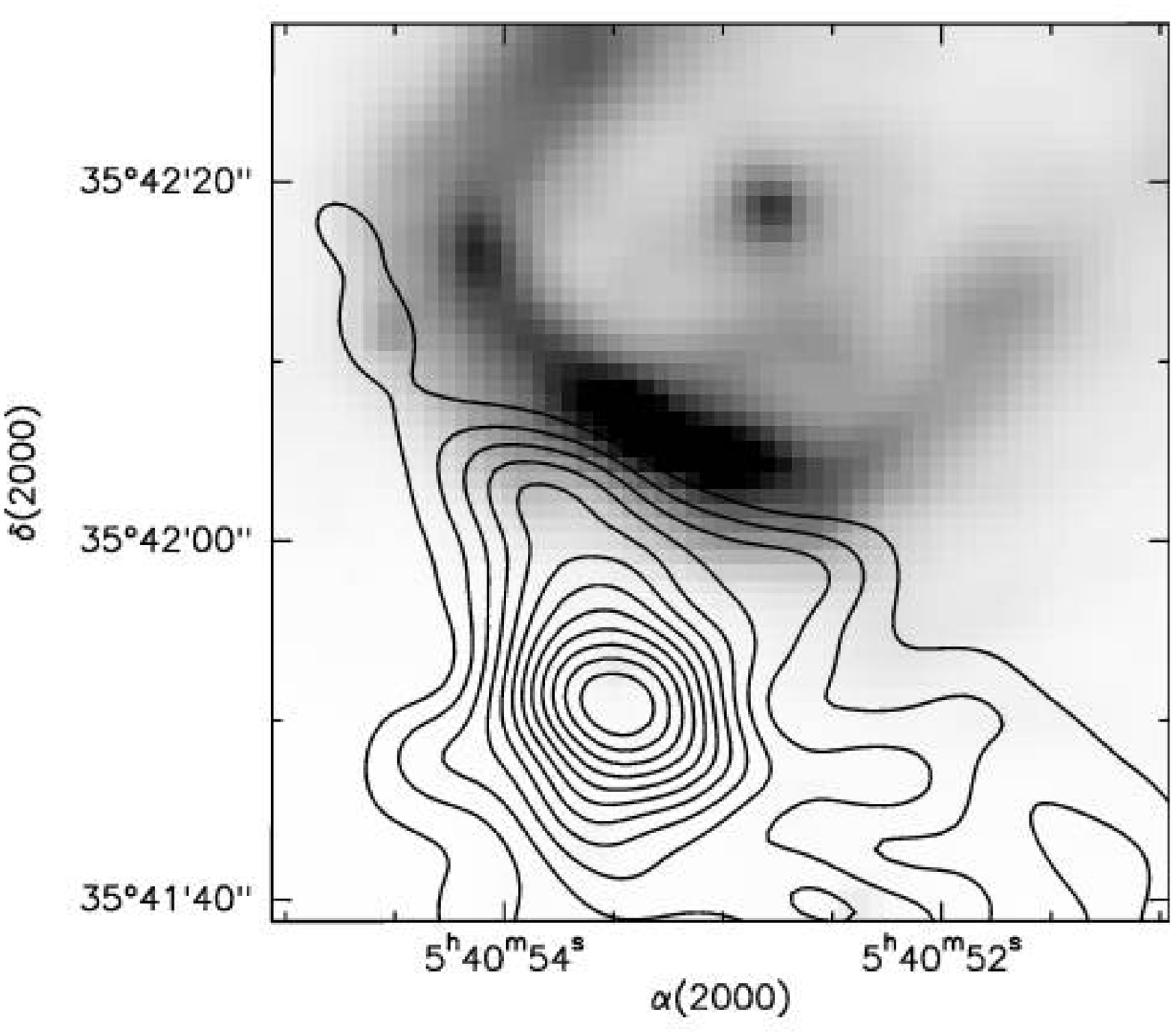}
      \caption{ The HCO$^{+}$ cloud at $-17$  km s$^{-1}$ (contours)
overlaid on the 5.8  $\mu$m Spitzer-IRAC image (grey scale). Note that 
the lower contours of the molecular emission closely follow
the boundary of the IR emission, suggesting that the S235A HII
region is interacting with the molecular cloud.
         \label{s235Amol}}
   \end{figure}

In Fig.~\ref{s235Amol}, we show the overlay of the HCO$^{+}$ integrated 
emission around  $-17$~km
s$^{-1}$ with the 5.8 $\mu$m Spitzer-IRAC image.  
The overlay indicates that the lower
contours of the molecular emission closely follow the outer boundary  of
S235A, suggesting that the HII region and the molecular cloud are
interacting. It is also worth noting that the mm core is just outside the S235A
boundary. This situation resembles that of some well-known cometary-shaped
UCHII regions such as G29.96--0.02 and G34.26+0.15 (Reid \& Ho 1985;
Wood \& Churchwell 1989; Fey et al. 1995), which face a density peak of
the molecular clump enshrouding them (Maxia et al. 2001; Gibb et al. 2004;
Watt \& Mundy 1999). In a number of cases, it has been found that such
a peak coincides with a hot molecular core, where
massive star formation is going on (Cesaroni et al. 1998;
Heaton et al. 1989; Garay \& Rodr\'{\i}guez 1990). In our case, the
temperature of the molecular core is 
%
$\sim30$ K (see Felli et al. 2004), well below the typical temperature
of hot molecular cores (see Kurtz et al.\ 2000),
and the HII region S235A has an asymmetric rather than a cometary shape; 
nevertheless, the interaction
between the ionized gas and the molecular cloud, as traced by
the overall structure
of the region, suggests that in this case also, one is observing an active
burst of star formation where different evolutionary phases (from cores to
evolved HII regions) co-exist. Whether the star formation episode in
the molecular core has been triggered by the expansion of the
nearby HII region S235A remains an open issue.

\subsection{VLA-1 and the jet}

\begin{figure}
   \centering
   \includegraphics[angle=-90,width=8cm]{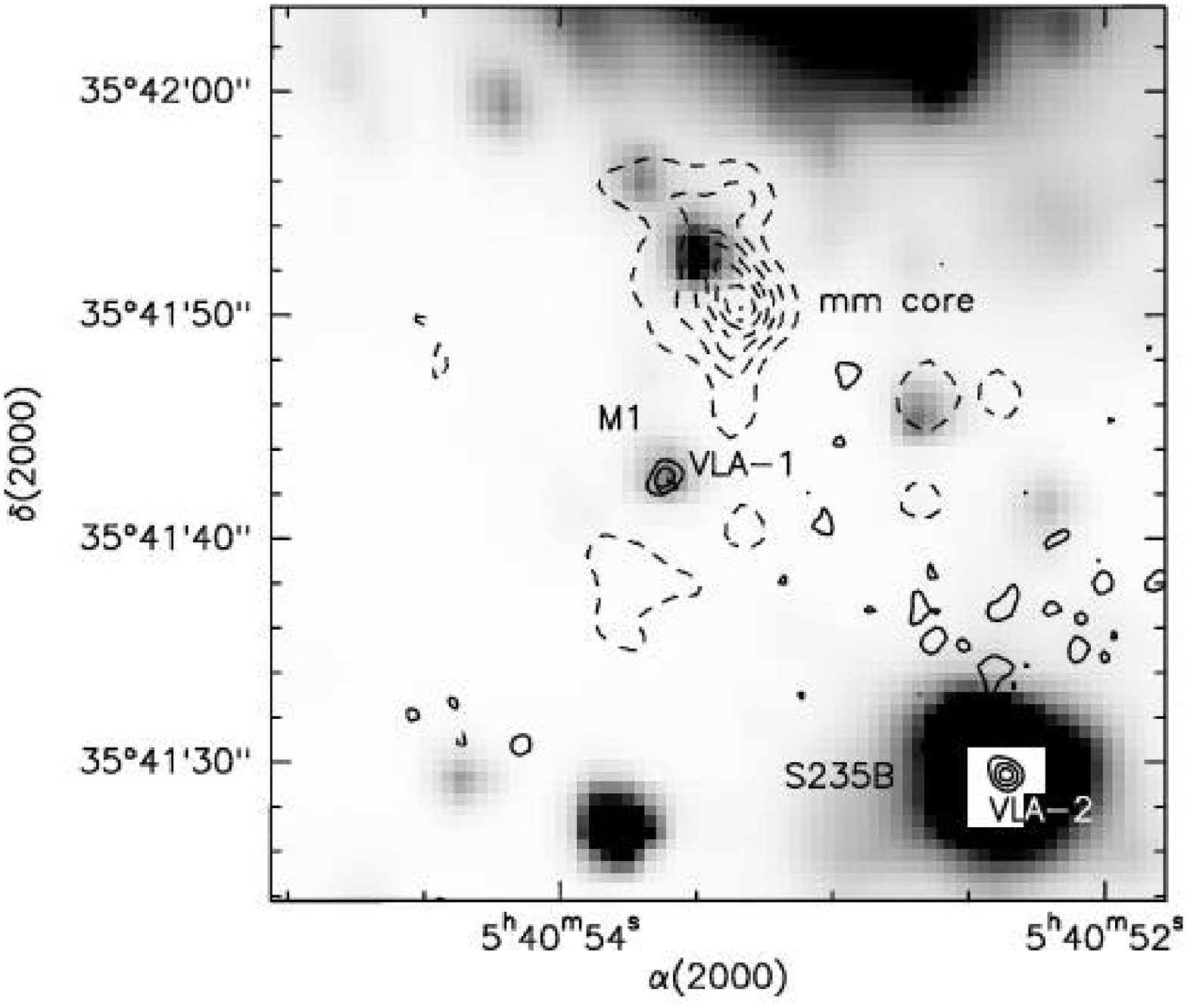}
    \caption{Overlay of the 1.3 cm VLA map (full contours)  with the 
Spitzer-IRAC 3.6 $\mu$m image (grey scale) and the 1.2 mm 
continuum (dashed contours).
         \label{spitzer11cm}}
   \end{figure}

Our radio continuum observations were unable  
to detect the elongated structure that had previously been found
at 3.3 mm, which also coincides with the blue lobe
of the NNW-SSE HCO$^{+}$(1--0) outflow. Instead,
a compact radio source, VLA-1, was found coincident with 
a small blob of emission in the central part of the 1.2 mm ``jet'', 
close (3$\arcsec$ east) to the weaker molecular peak at $-19$ km s$^{-1}$, 
called C19
(Felli et al.\ 2004). More importantly, VLA-1 coincides with the newly found
water maser S235A-B-H2O/1 and with M1, as shown in 
Fig.~\ref{spitzer11cm}, where the 1.3 cm VLA map is  
overlaid on the Spitzer-IRAC 3.6 $\mu$m image. 
M1, which had been previously assumed to be associated with
the $-60$ km s$^{-1}$ water maser (Felli et al.\ 1997), 
is instead associated with a a radio continuum source and a new, separate
water maser at $\sim -30$ km s$^{-1}$.

Although the errors on the flux densities are large, the values given 
in Table~\ref{cont:tab} for VLA-1 are consistent with
a spectral index $\alpha \approx 0.2$. This value is 
typical of partially optically thick free-free emission.

Since the present morphology no longer supports an ionized 
jet interpretation, we have to consider  the alternative 
possibility  that VLA-1 is an independent UCHII region, a hypothesis
that should also be considered, in light 
of the precise association of the water maser with M1. A lower limit
to the total number $N_{\rm Ly}$ of ionizing photons can be obtained from the
1.3 cm radio flux (the one with the highest signal-to-noise ratio) by
assuming that the HII regions are optically thin (Mezger 1978). We
obtain $N_{\rm Ly} \sim10^{44}$ s$^{-1}$, typical of B2-B3 ZAMS stars
(Panagia 1973).

What may have caused the apparent disagreement between the cm radio observations
which postulate the presence of a UCHII region and the mm observation that  
had suggested a jet?
Extrapolating the radio flux of VLA-1 according to $\nu^{\alpha}$
($\alpha \approx 0.2$), we derive $\sim0.8$ mJy
at 1.2 mm and $\sim0.6$ mJy at 3.3 mm. The 1.2 mm flux is below
$1\sigma$ ($\sim1$ mJy) in the Plateau de Bure map, so it could
not have been  detected as a point source. The 1.2 mm
flux of the ``jet'', 13 mJy, was integrated over a much larger 
area  defined by the 3.3 mm map.
At 3.3 mm, the extrapolated flux of VLA-1 is at a $2 \sigma$ level, 
again barely detectable as a point source, and in any case 
difficult to see since it would lie within the elongated 3.3 mm structure
(about $10\arcsec$ in the elongated direction and unresolved in the perpendicular
direction), with 7 mJy of integrated flux. 

These contradictory aspects can be reconciled if  
the emission from the elongated 3.3 mm structure comes 
from dust, perhaps that associated with 
the NNW-SSE blue outflow lobe or with  the C19 molecular peak. 
The spectral index $\alpha=0.6$  between 3.3 and 1.2 mm could be the result
of the simultaneous presence of a dusty jet with a steep spectral index
and a weak UCHII region showing up only at lower frequencies, where
dust emission is negligible.

Is VLA-1/M1
the driving source of the NNW-SSE outflow? Due to the overlap
with the NE-SW outflow, the centre of the NNW-SSE outflow
is ill-defined and any argument based on the position of
VLA-1 with respect to the centre of the NNW-SSE outflow is inconclusive.
However, we can exclude any association of VLA-1 with the NE-SW outflow,
given its offset with respect to its axis.

\begin{figure}
   \centering
   \includegraphics[angle=-90,width=8cm]{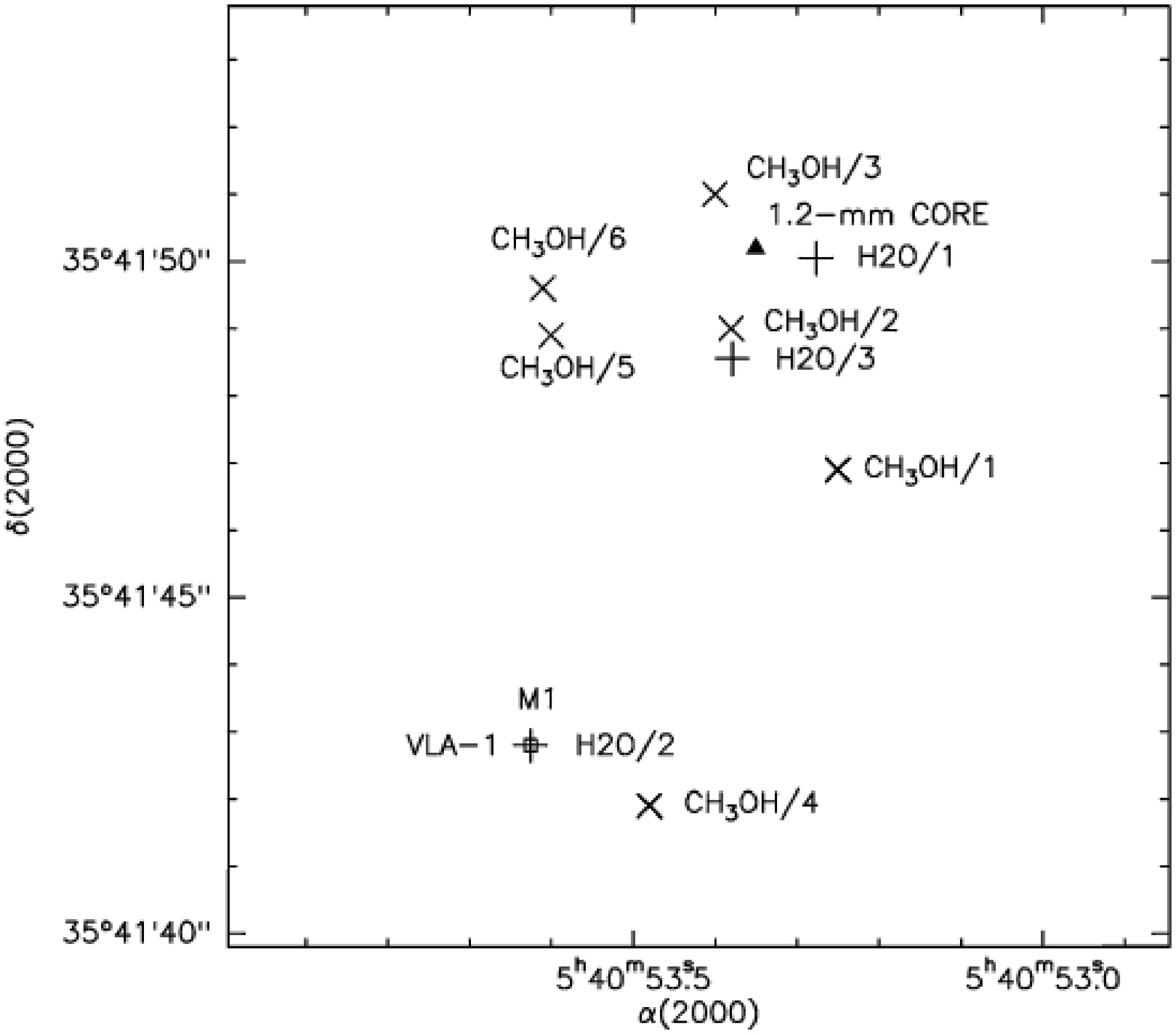}
    \caption{Position of the three water masers (H20/1-3 $+$) and of 
the six methanol masers (CH3OH/1-6 $\times$) detected by  Kurtz et al.\ (2004).
The 1.2 mm core 
(triangle), VLA-1 (square), and M1 are also indicated. Note that 
the symbol used to mark the location of the 1.2 mm core
is smaller than the errorbars.
         \label{Figmaser}}
   \end{figure}

Finally, we note that one 
of the methanol masers (CH$_3$OH/4) lies close (2$\arcsec$ 
southwest) to VLA-1 and the water maser 
S235A-B-H2O/2 (see Fig.~\ref{Figmaser})  and  has velocities in a 
similar range (from
$-16$ to $-21$ km s$^{-1}$, see Kurtz et al.\ 2004), suggesting a common origin.
The velocities of the water and methanol masers
are very close to that of the molecular core, indicating that in this 
case, the component of the motions along the line of sight
of the masers with respect to the molecular core is negligible.
%

%

\subsection{VLA-2 and S235B}

The overlay of Fig.~\ref{spitzer11cm} shows that VLA-2 lies at the centre
of the S235B bright diffuse nebula (saturated in all IRAC bands). 
The Spitzer-IRAC images, in particular 
those at longer wavelengths, 
%
indicate that this nebula is composed of two very close components
(see Fig.~\ref{spitzercolor}).
VLA-2 lies at the centre  of the southern  and more extended one. 
Overall, the size of the mid-IR nebula is  about 10$\arcsec$, similar to that
observed in H$\alpha$,  so that the optical-IR 
morphology is more reminiscent of S235A (resolved HII region) 
than that of VLA-1 (IR  and radio unresolved).

Besides being detected in H$\alpha$, strong Br$\gamma$ emission from
an unresolved source (i.e. $<$ 3$\arcsec$)  coincident with the
$K$-band point source in S235B had been reported
by Krassner et al.\ (1982) and Felli et al.\ (1997), with an
integrated line 
%
flux
of $F$(Br$\gamma$) = (2.0 $\pm$ 0.4) x 10$^{-12}$
erg s$^{-1}$ cm$^{-2}$.
The expected radio flux 
%
density
from an optically thin HII region at 6 cm
would have been  greater than 200 mJy, where the lower limit accounts 
for the fact that the 
Br$\gamma$ flux was not corrected for extinction.  This is at odds 
with lower limits to the radio flux found in all previous
radio continuum observations, as well as with the present detection. 
In the past, this forced 
%
abandon of
the hypothesis of a classical HII region (unless
extremely optically thick) 
%
suggested
that the Br$\gamma$
emission originates from an ionized expanding envelope around an early type
star (Felli et al. \ 1997), in which case the ratio of radio-to-IR line 
emission 
%
would be
reduced by about two orders of magnitude (Simon et al.\ 1983).

Our measured fluxes represent the first detection of the radio
continuum from an underlying unresolved source. 
The radio flux density at 3.6 cm
expected from a fully  ionized envelope  using
$F$(Br$\alpha$)/$F$(Br$\gamma$) $\sim0.8$
and Eq. (20) of Simon et al. (1983)
is 0.46 mJy. While this agreement with the observed value
might be fortuitous in view of the many unknown parameters involved 
(velocity of the 
wind, correction for extinction, etc.) it clearly points out that an
ionized envelope around a lower luminosity,more-evolved  star remains 
the best interpretation. However, it must be noted that
the observed spectral index is smaller than the expected 0.6 value. 

Our flux densities are fully
consistent with the previous upper limits at radio wavelengths and with
the unresolved nature of the 
%
S235B.
The implied mass loss using Eq. (14) of 
Felli \& Panagia  (1981) is $4\times10^{-6}$ M$_{\sun}$ yr$^{-1}$, quite large
and indicative of a luminous star.
Future radio recombination line observations
with sufficiently high sensitivity may 
%
permit measurement of
the line width to determine the velocity of the ionized wind.
There are no masers, molecular peaks,  outflows, or mm
peaks associated with S235B, all of which suggest 
%
that S235B-VLA-2 is more evolved than the mm core and VLA-1.

%
The diffuse H$\alpha$ and IR emission from 
S235B may be attributed to reflected light from ionized stellar envelopes.

Finally, 
%
what
is the luminosity and mass  of the star
embedded in  S235B? We cannot use 
%
the Spitzer-IRAC observations because they are saturated.
Instead, we used 
the J, H, and K magnitudes from Felli et al.\ (1997) and the MSX fluxes
(see Egan et al.\ 1999).
The integral of these values gives 410 L$_{\sun}$,
which must  be considered to be a  lower limit
%
because the FIR part of the spectrum is not taken into
account in our calculation.

   \begin{figure}
   \centering
   \includegraphics[width=8cm]{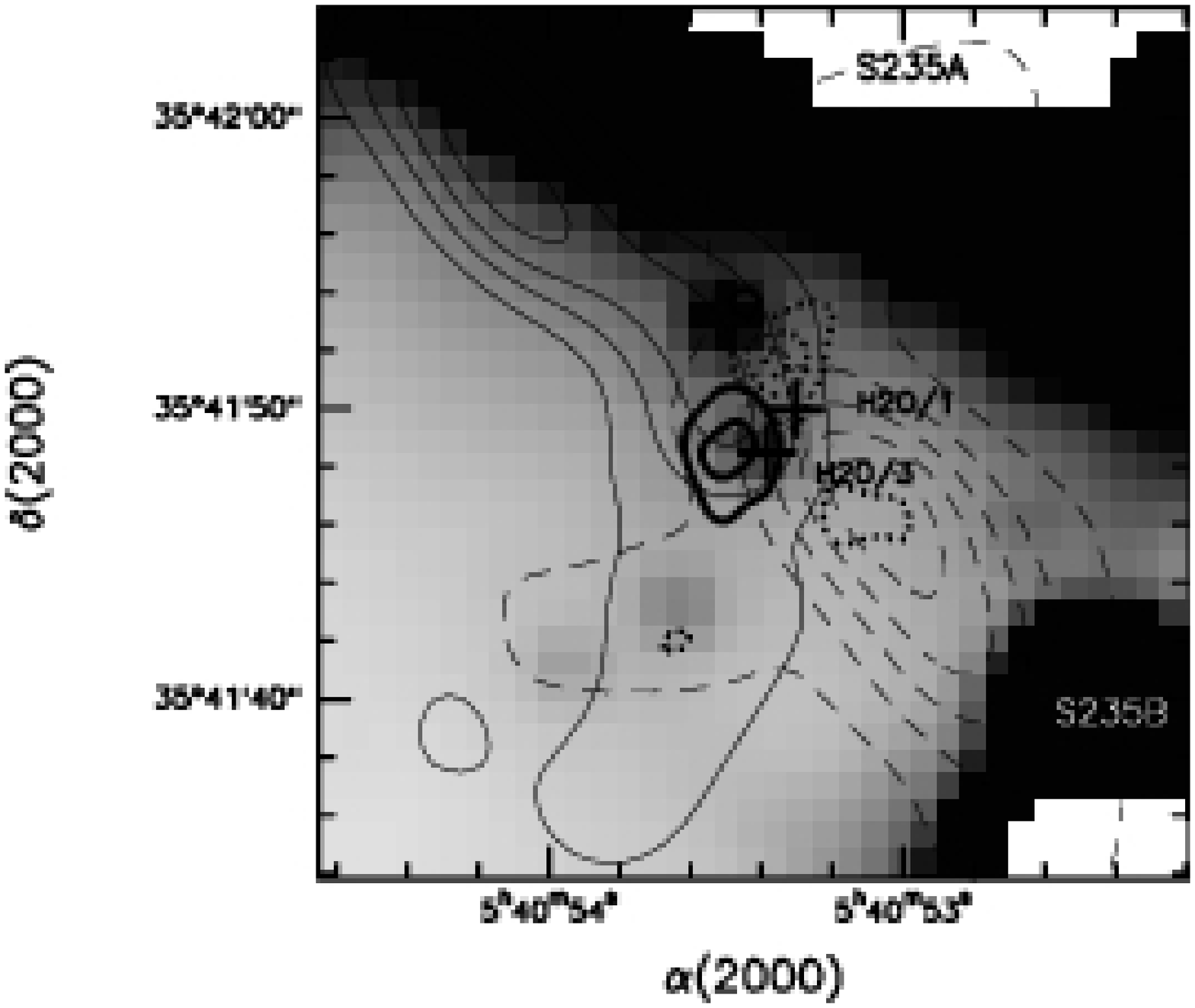}
      \caption{HCO$^{+}$(1--0) emission from the NE-SW outflow (thin contours)
      overlaid with the blue- and red-shifted emission of
      C$^{34}$S(5--4) (thick contours) from Felli et al.\ (2004). 
The ``$+$'' symbols  mark
      the location of the water maser spots H2O/1 ($\sim7$ km s$^{-1}$) 
and H2O/3 ($\sim-60$ km s$^{-1}$).
The background is the 8.0 $\mu$m  Spitzer-IRAC image (grey scale) and shows the
position of S235AB-MIR, coincident with the southern water maser
(H2O/3).
         \label{Fig3}}
   \end{figure}
%

\subsection{The mm core and the S235A-B-H2O/1 and S235A-B-H2O/3 masers}

As was already pointed out, an important result of the cm radio continuum
observations is the lack of emission 
%
from the mm core.
In Felli et al.\ (2004), the luminosity of the embedded YSO had
been estimated to be $10^3$ L$_{\sun}$.
If this comes from  a ZAMS star of spectral type B3 or earlier, 
to make the radio 
%
free-free
emission undetectable, the
radius of the associated HII region must be  less than $\sim$6.5~AU due to
confinement by a density larger than $7\times10^6$~cm$^{-3}$. In any case, the
detection of pure thermal dust emission from the core indicates a very early
evolutionary phase of the embedded YSO.

The new result provided by the Spitzer-IRAC observations is the 
detection of S235AB-MIR very close to the mm peak. 
The positions of the mm core and 
S235AB-MIR differ by $\sim1\farcs 5$, which is slightly greater
than the error on the relative positions, so that at this stage 
it cannot be firmly established if
S235AB-MIR represents the YSO embedded in the mm core. 
Its position in the colour-colour plots of Fig.~\ref{colcol} definitely puts 
this source in the Class I category.  

It is possible to check whether the S235AB-MIR fluxes
between $3.6$--8 $\mu$m are  consistent with the spectral energy distribution 
(SED) of a heavily embedded
(proto-)star of $10^3$ L$_{\sun}$. 
Felli et al.\ (2004) show that the non-detection of such a source in
the $K$ band towards the mm core implies $A_{V} \ga 37$
mag, in agreement with the derived H$_2$ column density. Using the extinction
law found by Indebetouw et al.\ (2005), we derived upper limits
for the intrinsic fluxes (or upper limits) of S235AB-MIR in the four 
IRAC bands. 
Of course, S235AB-MIR 
%
being
a Class I source, most of the emission at these wavelengths
arises from circumstellar matter rather than from the (proto-)star
photosphere. For this reason, we compared our results to the SEDs
models  for intermediate-mass Class I sources (star plus disk and envelope,  
Whitney et al.\ 2003;  Whitney et al.\ 2004). 
We found that, within the limits of 
the many parameters of their  
%
models,
the  MIR SED of
S235AB-MIR is consistent with a central star later than B3. 

The two northernmost water masers, S235A-B-H2O/1 and S235A-B-H2O/3, 
are located close to the mm core and  emit over velocity ranges 
quite different from
those of the other water and methanol masers, 
as well as those of the molecular
cloud ($\sim-17$ km s$^{-1}$). 
Figure~\ref{Fig3}  (adapted
from Fig.~21 of Felli et al.\ 2004)  shows the NE-SW HCO$^{+}$ outflow 
with the contours of a perpendicular bipolar
structure traced by the wings 
of C$^{34}$S(5--4). 
This was interpreted by Felli et al.\ (2004) as the signature of a rotating
disk around the YSO driving the outflow. Strikingly enough, the
two water masers are aligned in the same direction and very close to the 
C$^{34}$S(5--4) structures: the blue-shifted
maser (S235A-B-H2O/3 at $\sim -60$ km s$^{-1}$) lying towards 
the blue C$^{34}$S(5--4) lobe
(integrated from $-21$ to $-19$ km s$^{-1}$) 
and the red-shifted maser (S235A-B-H2O/1 at $\sim7$ km s$^{-1}$)
lying towards the red C$^{34}$S(5--4) lobe (from $-16$ to $-14$ km s$^{-1}$).
However, in both cases the velocity of the water masers are  more red or blue
shifted than the corresponding C$^{34}$S(5--4) lobes.

A simple calculation based on the maser  velocities 
shows that they cannot simply belong to 
a disk in Keplerian rotation around the YSO. From the difference between
the most extreme maser velocities (7 and $-68$ km s$^{-1}$), 
we can infer a lower limit to the rotation velocity  of $37$ km s$^{-1}$. 
If one assumes the half-distance between the two maser 
spots ($\sim1\arcsec$ or 1800 AU) to be the orbital radius, then  the mass needed
to maintain such a rotating disk should be $> 2500$ M$_{\sun}$,
much larger than the mass of the molecular core.

This fact proves that the water masers cannot trace rotation about an
embedded YSO.  One possibility is that they are instead the signature of the
interaction with the C$^{34}$S disk of high 
%
velocity outflowing material
from the YSO. The higher (red and blue) velocities of the water maser with
respect to C$^{34}$S and their closer position to the mm core may suggest
that the acceleration of the outflowing material occurs in the immediate
surroundings of the YSO.
 
Another possibility is that the C$^{34}$S emission is not tracing a disk,
but rather an outflow, whose high-velocity component would be seen in the
H$_2$O maser lines. This scenario is more consistent with the common belief
that water masers are strictly associated with jets powering molecular outflows
(Felli et al. 1992), as observed, e.g., in the massive protostar
IRAS\,20126+4104 (Moscadelli et al. 2000, 2005). If this hypothesis is
correct, it remains to be established whether the bipolar outflow seen in the
C$^{34}$S and H$_2$O lines would be the same as the NNW-SSE outflow or a
distinct one oriented approximately in the same direction, but originating
from a different YSO.
 
Of the two hypotheses presented above, we believe that the outflow origin for
the C$^{34}$S and H$_2$O emission is the most likely, given the tight
association between outflows and water masers. However, at present it is
impossible to rule out the possibility that the C$^{34}$S emission is coming
from a disk, as in the case of the Keplerian disk in IRAS\,20126+4104
(Cesaroni et al. 2005).  Class~I methanol masers are believed to be excited
in jets, so that a priori the 7~mm CH$_{3}$OH masers imaged by Kurtz et al.
(2004) in S235 could be used to establish the direction of the outflow and
hence choose between the two hypotheses.  As shown in
Fig.~\ref{Figmaser}, five of the maser spots cluster around the mm core
suggesting that for them, too, the main source of energy is the YSO within the
mm core. In particular, CH$_{3}$OH/2 is very close to S235A-B-H2O/3.  However
their velocities lie within a narrow range (from --15.9 to
$-21.0$~km~s$^{-1}$) and the spots do not show a clear bipolarity, either in
velocity or in their distribution.  It is therefore difficult to associate the
methanol maser emission to any precise geometry. The only conclusion is
that they are unlikely to have the same dynamical origin as the two water
masers.

\subsection{Water maser variability}

Some indication of what occurs in the immediate surroundings of the
YSO embedded within the mm core may come from the variability
of the two associated water masers. 

The emission at $\sim-60$ km s$^{-1}$
(S235A-B-H2O/3) reached its maximum (up to $\sim110$ Jy)
in 1992--1993, lasting at most
2 years, then disappearing for most of the Medicina  patrol, and finally
reappearing just before the VLA observations.

\begin{figure}
   \centering
   \includegraphics[angle=-90,width=8.5cm]{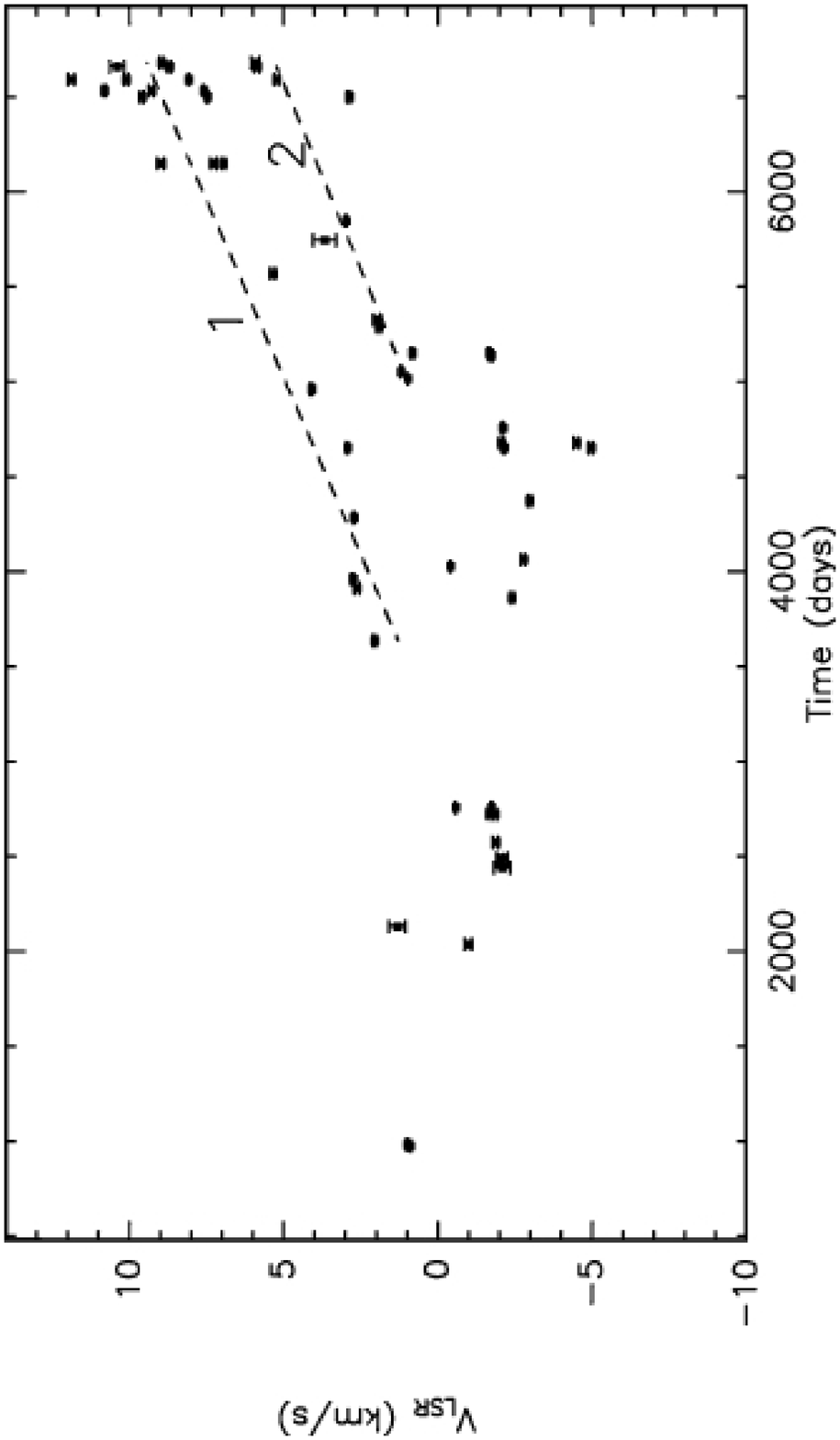}
      \caption{
%
      Velocities of the peaks of S235A-B-H2O/1
      from Gaussian fits to the data from
      the Medicina radio telescope. We have tentatively outlined,
      with dashed lines, two components (labelled as 1 and 2)
       whose velocities might be drifting during the period
      of our observations.
         \label{medi:maser:velo}}
   \end{figure}

Most noteworthy  is 
the emission from S235A-B-H2O/1 (between $-10$ and $10$ km s$^{-1}$), not only 
because of its long lasting presence and high variability,  but also
because of its velocity shifts,  $\pm 5$ km s$^{-1}$.
Figure~\ref{medi:maser:velo} shows the LSR  velocities 
of S235A-B-H2O/1 (obtained by Gaussian fitting)
in the period 1989-2005. 

While it is  possible  that the velocity variations are  simply due to
the random flaring of spots at different velocities, we shall investigate
more physically appealing interpretations.

To check whether the effect is due to a rotational modulation, we
tried to fit all the observed velocities (the source is above the noise
for about 60\% of the time)  with a sine function, as would be 
expected for a maser spot on a rotating disk viewed edge-on. 
%
No significant evidence was found.

The velocities in Fig.~\ref{medi:maser:velo} are all redshifted 
with respect to the systemic velocity,
but they are not all at random and seem to fan out at late periods.
We identify two  groups of
points  (labelled as 1 and 2 in
Fig.~\ref{medi:maser:velo}), which exhibit linear drifts of velocity
away from the systemic value.
A linear fit (dashed lines in Fig.~\ref{medi:maser:velo}) provides 
velocity drifts  of 
0.93 and  0.98 km s$^{-1}$ yr$^{-1}$, respectively.
Velocity drifts of this amount have been observed in other water masers that are
stronger and less variable in intensity (Brand et al. \ 2003).
It is tempting to explain the velocity drifts 
with shocked material that is accelerated from  
a mean velocity $\sim0$ km s$^{-1}$ 
by mass outflow from a central YSO. 
The lifetime of the accelerated spots is  $\sim1000-2000$ days
($\sim3-6$ yrs) and could be  related to the duration of 
%
ejection events from
the YSO.

The maser outflows from S235A-B-H2O/1 clearly deserve further, proper motion
studies with VLBI techniques.

%

\section{
%
Summary and conclusions
}

\label{conclu}
%
   \begin{figure}
   \centering
   \includegraphics[width=8cm]{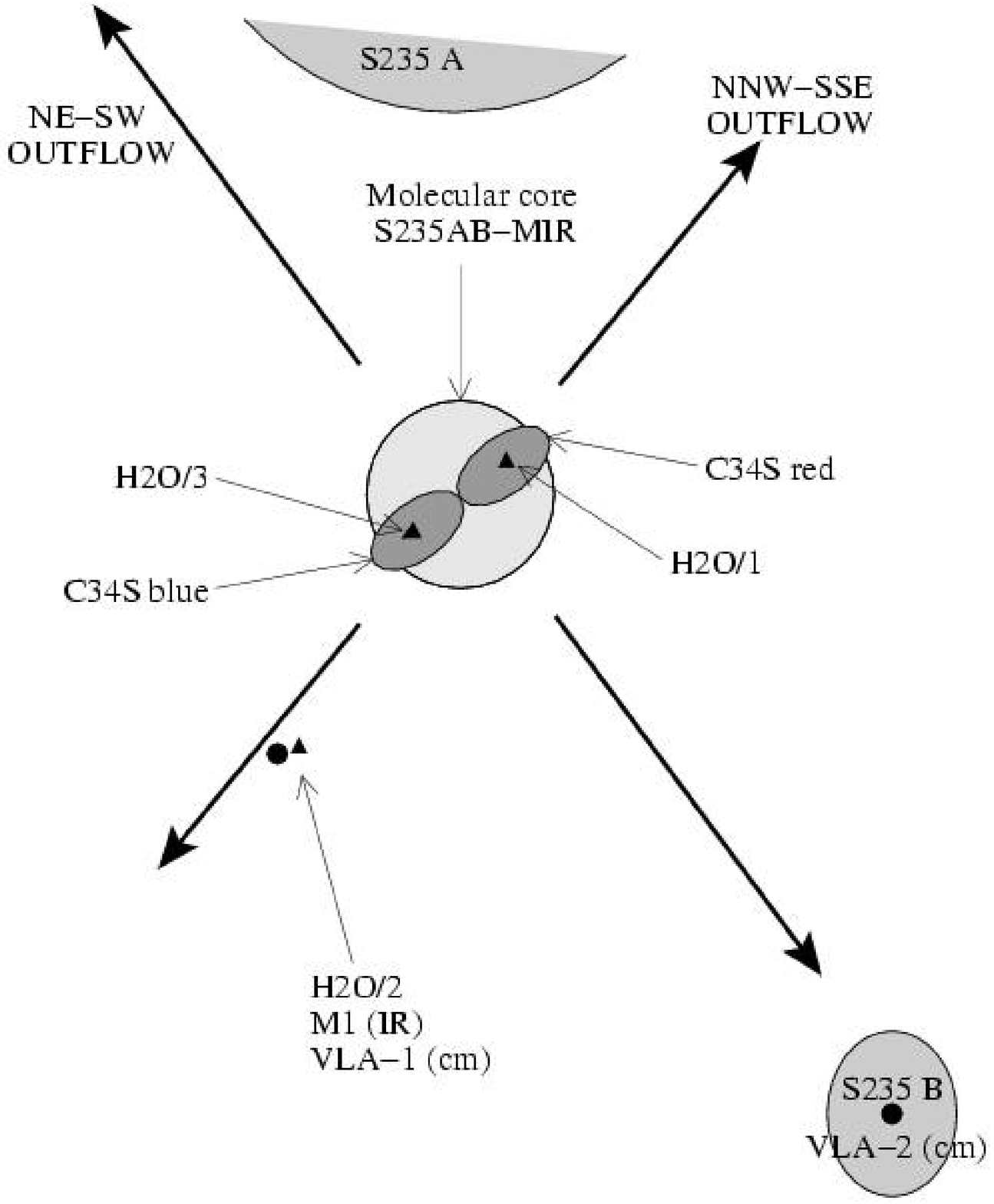}
      \caption{Sketch (not to scale)
of the star-forming region S235A-B in light
of the new data presented in this paper. New and already-known sources
are labelled, and their relationships are discussed in the text.
         \label{schema}}
   \end{figure}
We have presented new, more sensitive high-resolution VLA cm 
radio observations of the S235A-B region, as well as the results
of the Medicina  water maser patrol (started in 1987), and archive Spitzer-IRAC
observations. Several new aspects of this star-forming region emerge;
%
they are illustrated in Fig.~\ref{schema} and summarised in the
following:
   \begin{enumerate}
      \item
The radio-IR morphology of the  S235A HII region confirms that it is  
a classical HII region, optically thin in the cm range. 
It appears to interact with the molecular cloud and may
%
have induced
the formation of a second generation YSO in the mm core.
      \item 
No cm continuum emission is detected from the molecular core
discovered by Felli et al.\ (2004).
The lack of ionized hydrogen emission
suggests a very early evolutionary phase for the 
intermediate-luminosity embedded YSO,
much before the appearance of a UCHII region. 
We have found a new source, S235AB-MIR,
detectable only at 5.8 and 8.0 $\mu$m and close to the mm core, in the 
archival Spitzer images. 
Given its position in the colour-colour  plot, this could be 
the mid IR counterpart of the embedded YSO.
      \item 
We have observed no extended cm continuum emission from the 
elongated jet-like  structure  detected at 3.3 mm,
suggesting that the putative  3.3 mm ``jet'' is due
to dust and not to an ionized jet.
\item 
We found two compact radio-sources: VLA-1 and VLA-2. 
      Their spectral index is suggestive of partially thick free-free emission. 
      \item 
VLA-1 is located at the centre of
the elongated 3.3 mm structure and coincides
 with the near-IR source M1.  It is close to  the secondary molecular peak at 
$-19$ km s$^{-1}$.
We estimate that VLA-1  could be a UCHII region associated with a B2-B3 star.
\item 
We have discovered a water maser (S235A-B-H2O/2) at the same location of VLA-1.
A methanol maser (CH$_{3}$OH/4) lies close by at a similar velocity.
      \item  
VLA-2 is at the centre of S235B and represents the first radio
continuum detection from this source. Comparison with 
the near-IR hydrogen lines confirms that both emissions came from
an ionized envelope. 
      \item 
Two water masers (S235A-B-H2O/1 and S235A-B-H2O/3), 
      with very different velocities ($\sim7$ and $\sim-60$
      km s$^{-1}$, respectively), are located close to the mm core and
are aligned parallel to a structure found in C$^{34}$S and perpendicular to
      the NE-SW outflow. 
      \item 
Our single-dish observations with the Medicina radio telescope
do not spatially resolve the three water masers detected with the VLA,
although
%
do provide variability information
since they do not overlap in velocity.
\item 
A  high degree of variability in the water maser emission
was found in all cases. 
We found changes of the LSR  
      velocity with respect to the systemic velocity,
up to  $\sim5$ km s$^{-1}$ for S235A-B-H2O/1. 
A possible interpretation would be
velocity drifts due to
shocked gas accelerated by the flaring activity of the YSO.
\item 
The duration 
      of the acceleration, of an order of $3-6$ yrs, is similar to the
      lifetime of the emission.  
   \end{enumerate}

This paper, together with the preceding ones resulting from our long-term
study, reveals the simultaneous presence in the S235A-B complex of
widely different evolutionary phases, from the well-developed 
HII region S235A, to the peculiar object S235B-VLA-2, to the UCHII region
VLA-1, to the mm core which harbors a YSO in a very early  stage.
The comparison of high resolution multiwavelength observations, 
%
combined
with long-term monitoring of time variable phenomena,
provides unique information on the nature of young (proto)stars,
shedding new light on their interaction with the placental environment.
 
\begin{acknowledgements}

The Medicina observations are part of a long lasting project carried
out by the Arcetri-INAF and IRA-INAF water maser group (see e.g. Brand et
al.\ 2003 and references therein). 
This work is based in part on observations made with the Spitzer Space 
Telescope, which is operated by the Jet Propulsion Laboratory, California 
Institute of Technology under a contract with NASA.
This research made use of data products from the Midcourse Space Experiment
(MSX). Processing of the data was funded by the Ballistic Missile Defense 
Organization with additional support from the NASA Office of Space Science.
We acknowledge G. Comoretto and F. Palagi for their help in the study
of the variability of the water masers. 

\end{acknowledgements}

\end{document}